\documentclass[usenatbib]{mn2e}
\newcommand{\bib}{\bibitem[\protect\citeauthoryear}

\usepackage{graphicx}
\usepackage{epsf}
\usepackage{times}

\def\degrees{$^{\circ}$\ }
\def\srm{$\sigma_{\rm RM}~$}
\def\rmm{$\langle{\rm RM}\rangle~$}

\title[Faraday rotation measure bands across radio galaxies]
{
Ordered magnetic fields around radio galaxies: evidence for interaction
with the environment
}
\author[D. Guidetti et al.]
   {D. Guidetti\thanks{E-mail: dguidett@eso.org}$^{1,}$$^{2,}$$^{3}$, R.A. Laing $^1$, A.H. Bridle $^4$,
    P. Parma $^2$, L. Gregorini $^{3,}$$^{2}$\\
    $^1$ European Southern Observatory, Karl-Schwarzschild-Stra\ss e 2, D-85748 
    Garching-bei-M\"unchen, Germany \\ 
    $^2$ INAF -- Istituto di Radioastronomia, via Gobetti 101, I-40129 Bologna,
    Italy\\
    $^3$ Dipartimento di Astronomia, Univ. Bologna, via Ranzani 1, I--40127 Bologna, Italy\\
    $^4$ National Radio Astronomy Observatory, Edgemont Road, Charlottesville,
    VA 22903-2475, U.S.A.}

\begin{document}

\date{Accepted. Received; in original form }

\pagerange{\pageref{firstpage}--\pageref{lastpage}} \pubyear{2002}

\maketitle

\label{firstpage}
\begin{abstract}

We present detailed imaging of Faraday rotation and depolarization for the radio
galaxies 0206+35, 3C\,270, 3C\,353 and M\,84, based on Very Large Array
observations at multiple frequencies in the range 1365 to 8440\,MHz.  All of the
sources show highly anisotropic banded rotation measure (RM) structures with contours of
constant RM perpendicular to the major axes of their radio lobes.  All except
M\,84 also have regions in which the RM fluctuations have lower amplitude and
appear isotropic.  We give a comprehensive description of the banded RM
phenomenon and present an initial attempt to interpret it as a consequence of
interactions between the sources and their surroundings.  We show that the
material responsible for the Faraday rotation is in front of the radio emission
and that the bands are likely to be caused by magnetized plasma which has been
compressed by the expanding radio lobes.  We present a simple model for the
compression of a uniformly-magnetized external medium and show that RM bands of
approximately the right amplitude can be produced, but for only for special
initial conditions. A two-dimensional magnetic structure in which the field
lines are a family of ellipses draped around the leading edge of the lobe can
produce RM bands in the correct orientation for any source orientation.  We also
report the first detections of rims of high depolarization at the edges of the
inner radio lobes of M\,84 and 3C\,270. These are spatially coincident with
shells of enhanced X-ray surface brightness, in which both the field strength
and the thermal gas density are likely to be increased by
compression. The fields must be tangled on small scales.
\end{abstract}


\section{Introduction}
\label{sec:intro}

The detection of diffuse synchrotron emission (radio halos) on Mpc scales in an
increasing number of galaxy clusters provides good evidence for a distributed
magnetic field of $\mu$Gauss strength in the hot intracluster medium (ICM; see
e.g. \citealt{Ferrari08} for a review).  Imaging of Faraday rotation of
linearly-polarized radio emission from embedded and background sources confirms
that there are fields associated with thermal plasma along lines of sight
through the clusters (e.g.\ \citealt{CT02}).
Observations of Faraday rotation can also be
made for radio galaxies in sparser environments, allowing the study of magnetic
fields in environments too sparse for radio halos to be detected (e.g.\
\citealt{laing08,Guidetti10}).

The Faraday effect \citep{Fara1846} is the rotation suffered by linearly polarized
 radiation travelling through a magnetized medium, 
and can be described by the two following relations:
\begin{equation}
\label{pang}
\Delta\Psi_{[{\rm rad}]}= \Psi(\lambda)_{[{\rm rad}]}~-~\Psi_{0~[{\rm
      rad}]}=\lambda^2_ {[{\rm m}^2]}~{\rm RM}_{[{\rm rad\,m}^{-2}]},
\end{equation}
with
\begin{equation}
{\rm RM}_{[{\rm rad\,m}^{-2}]}=812\int_{0}^{L_{[{\rm kpc}]}}n_{\rm{e}~[{\rm
      cm}^{-3}]}B_{z~[\mu{\rm G}]}dz_{[{\rm kpc}]}\,,
\label{equarm}
\end{equation}
where  $\Psi(\lambda)$ and $\Psi_0$ are the ${\bf E}$-vector position angle of linearly
polarized radiation observed at wavelength $\lambda$ and
the intrinsic angle, respectively, $n_e$ is the electron gas density, $B_{z}$ is the magnetic
field along the line-of-sight ($B_{\parallel}$), and 
$L$ is the integration path. RM is the {\em rotation measure}.

Observations of Faraday rotation variations across extended radio
galaxies allow us to derive information about the integral of the
density-weighted line-of-sight field component. The hot ($\rm T\simeq10^7-10^8$\,K)
plasma emits in the X-ray energy band via thermal bremsstrahlung. When high quality X-ray
data for a radio-source environment is available, it is possible to infer the gas
density distribution and therefore to separate it from that of the magnetic
field, subject to some assumptions about the relation of field strength and density.

\begin{figure*}
\centering
\includegraphics[width=18.5cm]{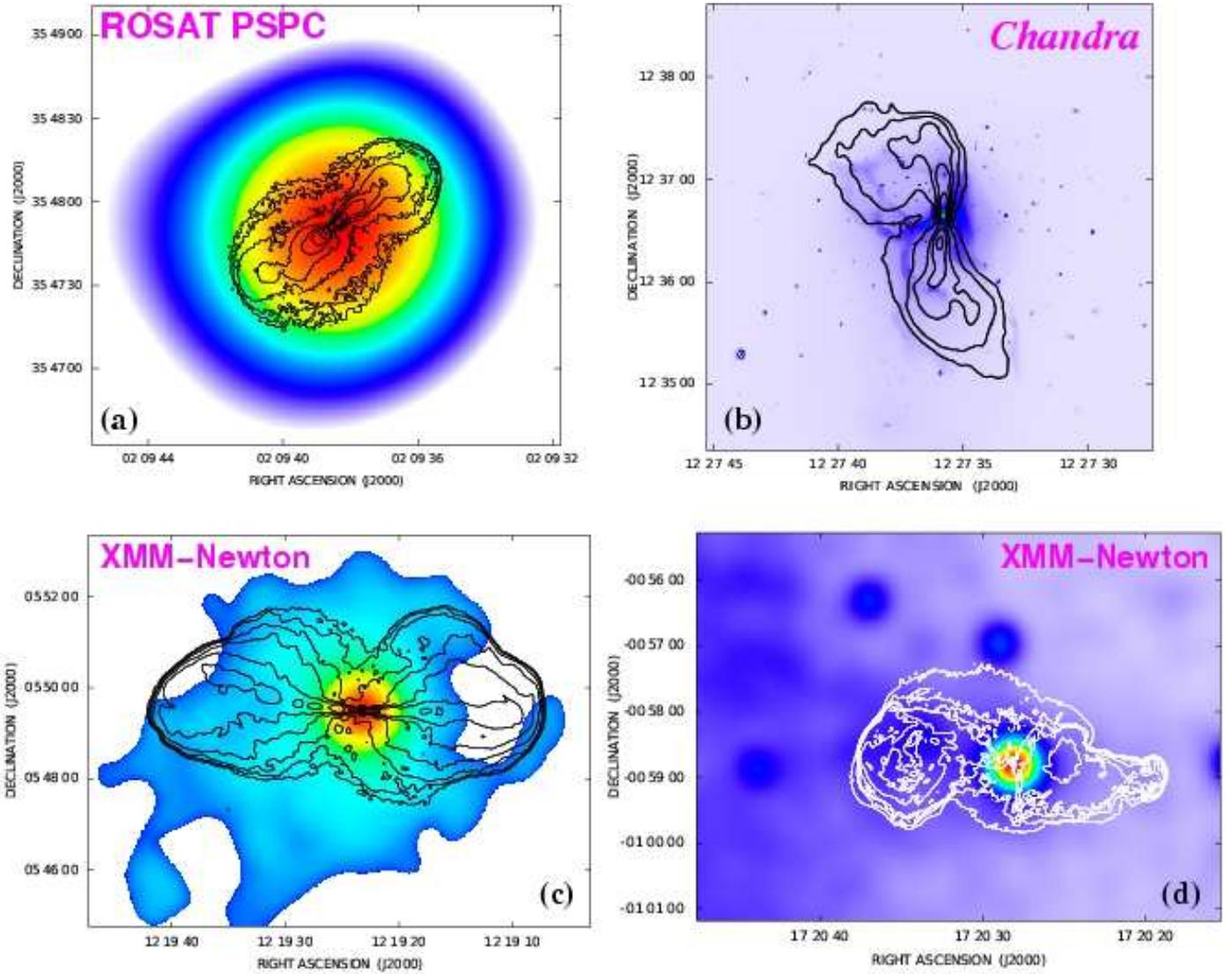}
\caption[]{
X-ray images overlaid with radio contours for all sources:
(a) 0206+35: 1385.1\,MHz VLA radio map with 1.2\,arcsec FWHM; the contours
are spaced by factor of 2 between 0.06 and 15\,mJy beam$^{-1}$. The ROSAT PSPC image
\citep{Worr00}, is smoothed with a Gaussian of $\sigma=30$\,arcsec.
(b) M\,84: 1413.0\,MHz VLA radio map with 4.5\,arcsec FWHM;
the contours are spaced by factor of 2 between 1 and 128\,mJy beam$^{-1}$.
The {\sl Chandra} \citep{Fino01} image is a wavelet reconstruction on angular scales
from 4 up to 32\,arcsec. 
(c) 3C\,270: 1365.0\,MHz VLA radio map with 5.0\,arcsec FWHM;
the contours are spaced by factor of 2 between 0.45 and 58\,mJy beam$^{-1}$. The XMM-Newton image
\citep{Cro05} is smoothed with a Gaussian of $\sigma=26$\,arcsec.
(d) 3C\,353: 1385.0\,MHz VLA radio map with 1.3\,arcsec FWHM;
the contours are spaced by factor of 2 between 0.35 and 22\,mJy beam$^{-1}$. The XMM-Newton image
\citep{Good08} is smoothed with a Gaussian of $\sigma=30$\,arcsec.
The {\sl Chandra} image of 3C\,270 is displayed in logarithmic scale.
}
\label{X}
\end{figure*}
Most of the RM images of radio galaxies published so far show patchy structures
with no clear preferred direction, consistent with isotropic foreground
fluctuations over a range of linear scales ranging from tens of kpc to
$\la$100\,pc (e.g.\ \citealt{Govoni06,Guidetti08,laing08,Guidetti10}).
Numerical modelling has demonstrated that this type of complex RM structure can
be accurately reproduced if the magnetic field is randomly variable with
fluctuations on a wide range of spatial scales, and is spread throughout the
whole group or cluster environment (e.g.\
\citealt{Murgia,Govoni06,Guidetti08,laing08,Guidetti10,Vacca10}).  These authors
used forward modelling, together with estimators of the spatial statistics of
the RM distributions (structure and autocorrelation functions or a multi-scale
statistic) to estimate the field strength, its relation to the gas density and
its power spectrum.  The technique of Bayesian maximum likelihood has also been
used for this purpose \citep{EV05,Kuchar09}.

In order to derive the three-dimensional magnetic field power spectrum, all of
these authors had to assume statistical isotropy for the field, since only the
component of the magnetic field along the line-of-sight contributes to the
observed RM.  This assumption is consistent with the absence of a 
preferred direction in most of the RM images.

In contrast, the present paper reports on \textit{anisotropic} RM structures,
observed in lobed radio galaxies located in different environments, ranging
from a small group to one of the richest clusters of galaxies.
The RM images of radio galaxies presented in this paper show clearly anisotropic
``banded'' patterns over part or all of their areas. In some sources, these banded patterns
coexist with regions of isotropic random variations. 
The magnetic field responsible for these RM patterns
must, therefore, have a preferred direction.

One source whose RM structure is dominated by bands is already known: M\,84
\citep{LB87}. In addition, there is some evidence for RM bands in sources which also show
strong irregular fluctuations, such as Cygnus\,A \citep{CT02}.  It is possible,
however, that some of the claimed bands could be due to imperfect sampling of an isotropic 
RM distribution with large-scale power, and we return to this question in Section~\ref{many}.

The present paper presents new 
RM images of three sources which show
spectacular banded structures,
together with improved data for M\,84. The
environments of all four sources are well characterized by modern X-ray
observations, and we give the first comprehensive description of the banded
RM phenomenon. We present an initial attempt to interpret
the phenomenon as a consequence of source-environment interactions and to
understand the difference between it and the more usual irregular RM structure. 

The RM images reported in this paper are derived from new or previously
unpublished archive  Very Large Array (VLA)\footnote{The Very Large Array is
a facility of the National Science Foundation, operated under 
cooperative agreement by Associated Universities, Inc.} data for the
nearby radio galaxies 
0206+35, M\,84 \citep{laing11}, 3C\,270 \citep{LB11a} and 3C\,353 (Swain, private
communication; see \citealt*{Swain96}).

The paper is organized as follows.
In Section\,\ref{sec:data}  the radio and X-ray properties of the 
sources under investigation are presented and in Section\,\ref{sec:teq} we briefly summarize the  techniques
used to analyse depolarization and two-dimensional variations of RM.
Sections\,\ref{sec:RM} and \ref{dp} present the RM and depolarization images
on which our analysis is based and correlations between the two quantities.
In Section\,\ref{sec:sfunc}, we evaluate the RM structure functions in regions
where the fluctuations appear to be isotropic and derive the power spectra.
A simple model of the source-environment interaction which characterises the effects
of compression of a magnetised IGM is described in Section~\ref{model}. This can
produce RM bands, but only under implausible special initial
conditions. Empirical ``draped'' field configurations which are  able to reproduce
the banded RM distributions  are investigated in Section\,\ref{drap}. 
In Section\,\ref{discuss}, we speculate on correlations between radio source morphology,
and RM anisotropy, discuss other examples from the literature, consider the effects of an
isotropic foreground Faraday screen on the detectability of RM bands and  briefly discuss 
possible asymmetries in the amplitude of the RM bands between
the approaching and receding lobes.
Finally, Section\,\ref{concl} summarizes our conclusions.

Throughout this paper we assume a $\Lambda$CDM cosmology with
$H_0$ = 71 km s$^{-1}$Mpc$^{-1}$,
$\Omega_m$ = 0.3, and $\Omega_{\Lambda}$ = 0.7.

\section{The Sample}
\label{sec:data}

High quality radio and X-ray data are available for all of the sources.
In this Section we  summarize those of their observational properties which are relevant to our RM study.
A list of the sources and their general parameters is given in Table\,\ref{propradio}, while
Table\,\ref{propX} shows the X-ray parameters taken from the literature and
equipartition parameters derived from our radio observations.

The sources were observed with the VLA at several frequencies, 
in full polarization mode and 
with multiple configurations so that the radio structure is well sampled.
The VLA observations, data reduction and detailed descriptions of the radio
structures are given for 0206+35 and M\,84 by  
\citet{laing11}, for 3C\,270 by \citet{LB11a}, and for 3C\,353
by \citet{Swain96}.
All of the radio maps 
show a core, two sided jets and a double-lobed structure
with sharp brightness gradients at the leading edges of both lobes. The
synchrotron minimum pressures are all significantly lower than the thermal
pressures of the external medium (Table\,\ref{propX}).

All of the sources have been observed in the soft X-ray band by more than one 
satellite, allowing the detection of multiple components on cluster/group and
sub-galactic scales. The X-ray morphologies are characterized by 
a compact source surrounded by extended emission with low surface brightness.
The former includes a non-thermal contribution, from the core and the inner 
regions of the radio jets and, in the case of 0206+35 and 3C\,270, a thermal
component which is well fitted by a small core radius $\beta$ model.
The latter component is associated with the diffuse intra-group or intra-cluster medium.
Parameters for all of the thermal components, derived from X-ray observations, are listed in Table\,\ref{propX}.
Because of the irregular morphology of the hot gas
surrounding 3C\,353 and M\,84, it has not been possible to fit $\beta$ models 
to their X-ray radial surface brightness profiles.

\begin{table*}
\caption{General optical and radio properties: Col. 1: source name; Cols. 2\&3: position; 
Col. 4: redshift; Col. 5: conversion from angular to spatial scale with the adopted cosmology; 
Col. 6: Fanaroff-Riley class; Col. 7: the largest angular size of the radio source; 
Col. 8: radio power at 1.4\,GHz; Col. 9:  angle to the line of sight of the jet axis;
Col. 10: radio spectral index; Col. 11: environment of the galaxy; Col. 12: reference.
}
\label{propradio}
\centering
\begin{tabular} {@{}c c c c c c c c c c c c}  
\hline
source & RA & DEC & z & kpc/arcsec & FR class & LAS & log${~P_{1.4}}$  &  $\theta $ & env. & ref.\\
       & [J2000] & [J2000] & & &  & [arcsec]  &  [W~Hz$^{-1}$] &  [degree] &  & &  \\
\hline
0206+35 (4C\,35.03) & 02 09 38.6 & +35 47 50 & 0.0377 & 0.739 & I & 90  & 24.8 &  40 & group & 1 \\
3C\,353 & 17 20 29.1 & -00 58 47 & 0.0304 & 0.601 & II  & 186  & 26.3 & 90  & poor cluster & 2   \\
3C\,270 & 12 19 23.2 & +05 49 31  & 0.0075 & 0.151 & I    & 580   & 24.4 & 90  &  group & 3 \\
M\,84   & 12 25 03.7 & +12 53 13 &  0.0036 & 0.072 & I & 150  & 23.2 & 60 & rich cluster & 3 \\
\hline
\multicolumn{12}{l}{\scriptsize References for the environmental classification: 
(1) \citet{Miller02}; (2) \citet{deV91}; (3) \citet{Trag00}.}
\end{tabular}

\end{table*}

\begin{table*}
\caption{X-ray and radio equipartition parameters for all the sources. Col. 1: source name;
Col. 2: X-ray energy band; Col. 3: average thermal temperature; 
Cols. 4,5,6 and 7,8,9 best-fitting core radii, central densities and $\beta$
parameters for the outer and inner $\beta$ models, respectively;
Col. 10: average thermal pressure at the midpoint of the radio lobes; Cols. 11\&12:
minimum synchrotron pressure and corresponding magnetic field; 
Col.13: references for the X-ray models.
}
\centering
\begin{tabular} {@{}c c c c c c c c c c c c c}  
\hline
source & band & $kT$ &  $r_{cx_{out}}$ & $n_{0_{out}}$ & $\beta_{out}$ & $r_{cx_{in}}$ &$n_{0_{in}}$ & $\beta_{in}$ &
 $P_0$ &  $P_{min}$ & $B_{\rm Pmin}$  &  ref.  \\
       & [keV] & [keV] & [kpc] & [cm$^{-3}$]& & [kpc] & [cm$^{-3}$] &  & [dyne~cm$^{-2}$] & [dyne~cm$^{-2}$] & $\mu$G & \\
\hline
0206+35 & 0.2-2.5 & 1.3$^{+1.3}_{-0.3}$  & 22.2 & 2.4~$\times$~10$^{-3}$ & 0.35 & 0.85 & 0.42 & 0.70 & 9.6~$\times$~10$^{-12}$  &  4.31$\times$~10$^{-13}$ & 5.70 & 1, 2 \\
3C\,353 & '' &    4.33$^{+0.25}_{-0.24}$ & &  & &  & &  &  & 1.66$\times$~10$^{-12}$ & 11.2 & 3  \\      
3C\,270 & 0.3-7.0 & 1.45$^{+0.23}_{-0.01}$  & 36.8 & 7.7~$\times$~10$^{-3}$ & 0.30 & 1.1 & 0.34 & 0.64 &  5.75$\times$~10$^{-12}$& 1.64$\times$~10$^{-13}$ & 3.71 & 4  \\         
M\,84   & 0.6-7.0 &  0.6$^{+0.05}_{-0.05}$ &   &    &  & 5.28$\pm$0.08  & 0.42  & 1.40$\pm$0.03    & 1.70$\times$~10$^{-11}$   & 1.07$\times$~10$^{-12}$  & 9.00 & 5     \\
\hline
\multicolumn{13}{l}{\scriptsize References: (1)
\citet{Worr00}; (2)\citet{Worr01}; (3) \citet{Iwa00}; (4) \citet{Cro08}; (5)
\citep{Fino01}.}
\label{propX}
\end{tabular}
\end{table*}

\subsection{0206+35 (4C\,35.03)}
\label{02}

0206+35 is an extended Fanaroff-Riley Class I (FR\,I; \citealt{Fana74}) radio source 
whose optical counterpart, UGC1\,1651, is a D-galaxy, a member of a dumb-bell system 
at the centre of a group of galaxies.
At a resolution of 1.2\,arcsec the radio emission shows a core, with smooth two-sided 
jets aligned in the NW-SE direction and surrounded by a diffuse and symmetric halo.
\citet{LB11b} have estimated that the jets are inclined by $\approx$40\degrees
with respect to the line of sight,  with the main (approaching) jet in the NW direction.

0206+35 has been observed with both the {\sl ROSAT} PSPC and HRI instruments \citep{Worr94,Worr00,Trus97}
and with {\sl Chandra} \citep*{Worr01}.
The X-ray emission consists of a compact source surrounded by a galactic 
atmosphere which merges into the much more extended intra-group gas.
The radius of the extended halo observed by the {\sl ROSAT} PSPC is $\approx$2.5\,arcmin (Fig. \,\ref{X}a).
The ROSAT and {\sl Chandra} X-ray surface brightness profiles are well fit by
the combination of $\beta$ models with two different core radii and a power-law
component (Hardcastle, private communication; Table\,\ref{propX}).

\begin{figure*}
\centering
\includegraphics[width=18.5cm]{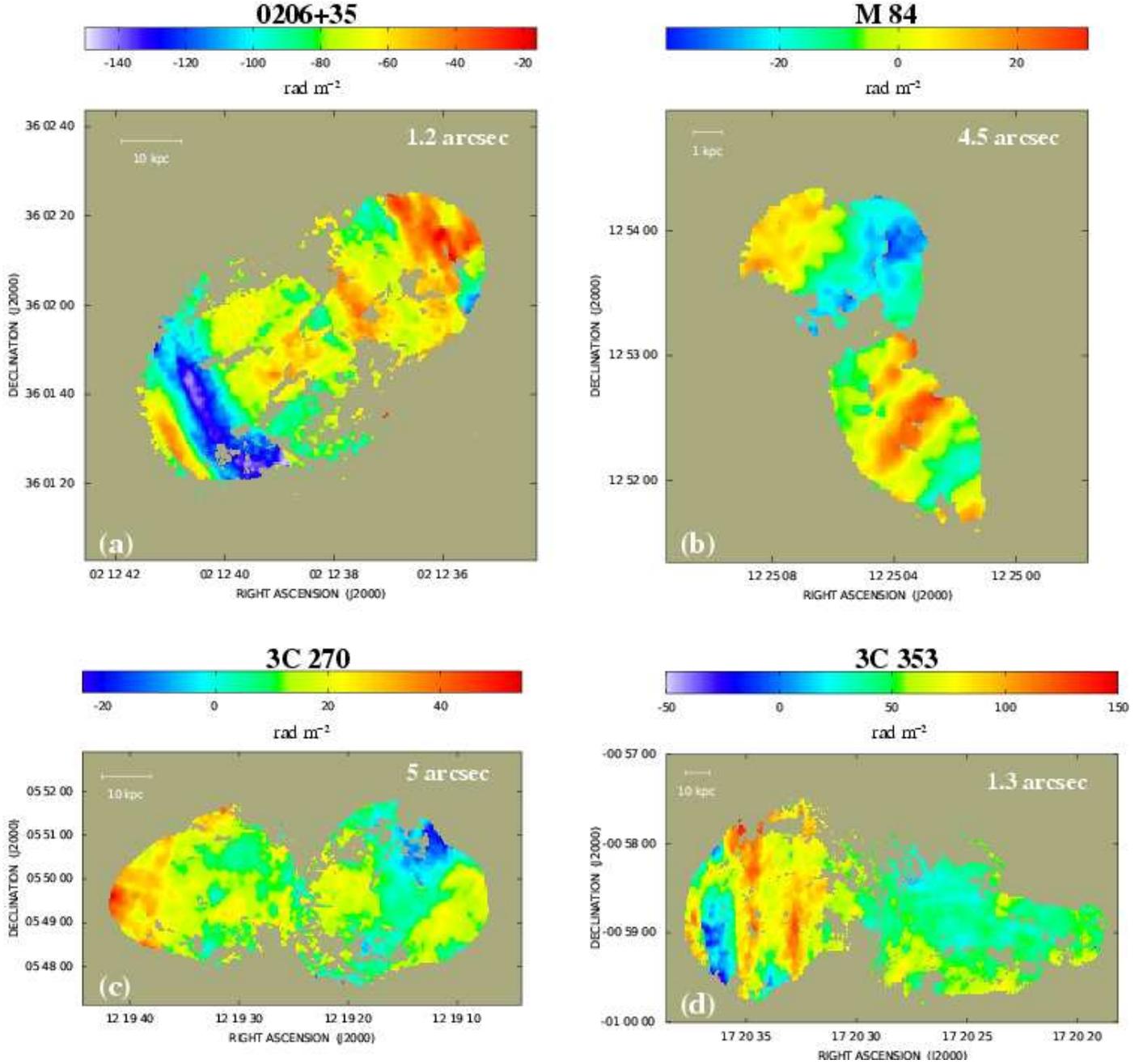}
\caption[]{RM images for all sources: (a) 0206+35; (b) M\,84; (c) 3C\,270; (d) 3C\,353.
The angular resolution and the linear scale of each map are shown in the individual panels. 
}
\label{RM}
\end{figure*}

\begin{figure*}
\centering
\includegraphics[width=18.3cm]{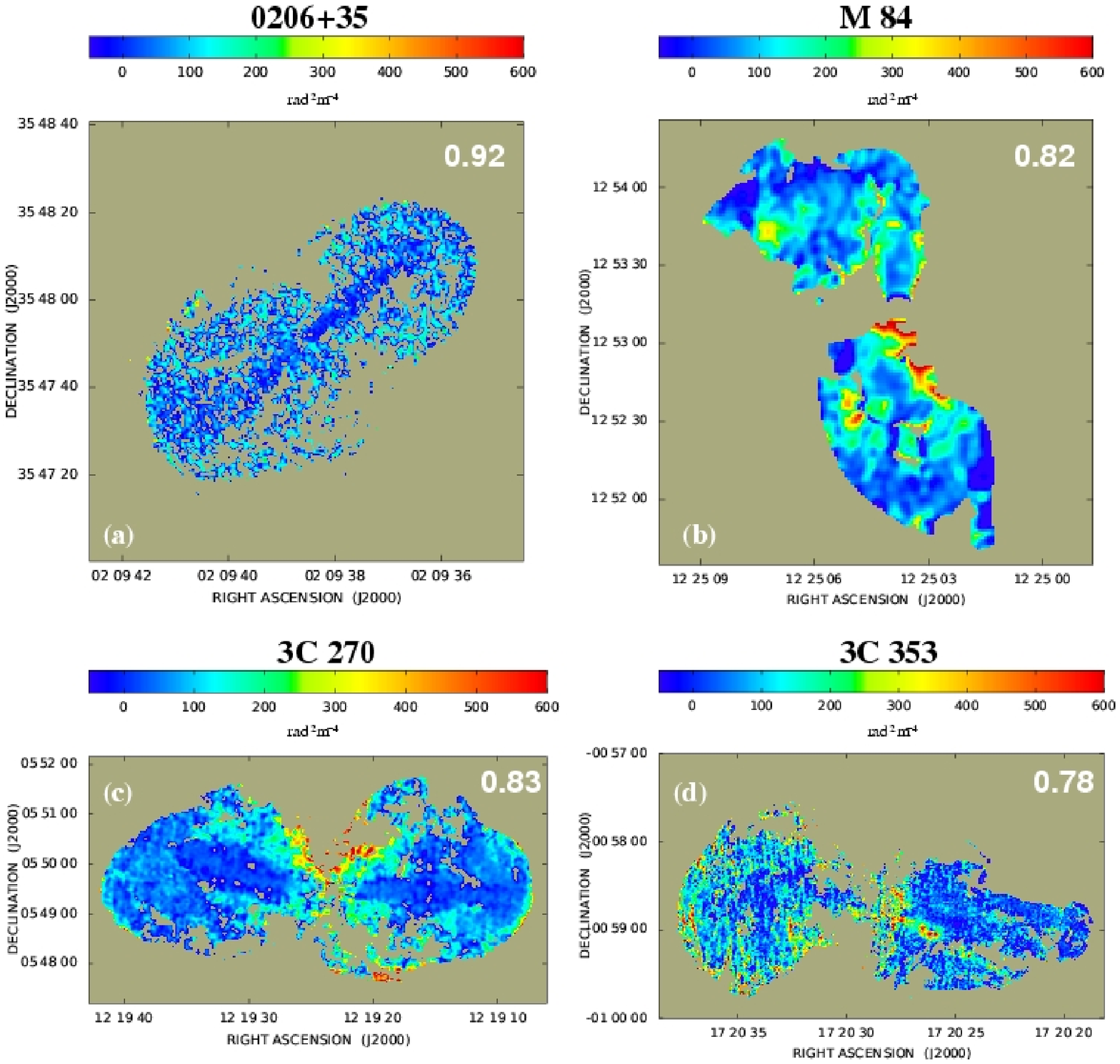}
\caption[]{Burn law k images for all sources at the same angular resolutions 
as for the RM images: (a) 0206+35 at 1.2\,arcsec FWHM; (b) M\,84 at
 4.5\,arcsec FWHM; (c) 3C\,270 at 5\,arcsec FWHM; (d) 3C\,353 at 1.3\,arcsec FWHM.
The corresponding integrated depolarization (DP; Section~\ref{dp}) is indicated on the top right angle of each panel.
The colour scale is the same for all displays.
}
\label{k}
\end{figure*}

\subsection{3C\,270}
\label{270}

3C\,270 is a radio source classified as FR\,I in most of the literature,
although  in fact, the two lobes have different FR classifications at low
resolution \citep{LB11a}.
The optical counterpart is the giant elliptical galaxy NGC\,4261,  
located at the centre of a nearby group. 
The radio source has a  symmetrical structure with a bright core and twin jets,
extending E-W and completely surrounded by lobes. 
The low jet/counter-jet ratio indicates that the jets are close to the plane of
the sky, with the Western side approaching \citep{LB11a}.

The XMM-Newton image (Fig.\,\ref{X}c) shows a disturbed distribution with regions
of low surface-brightness (cavities) 
at the positions of both radio lobes.
A recent {\sl Chandra} observation \citep{Worr10} shows ``wedges'' of low X-ray
surface brightness surrounding the inner jets  
(see also \citealt*{Cro05}, \citealt{Fino06}, \citealt{Jetha07}, \citealt{Cro08}).
The overall surface brightness profile is accurately reproduced by
a point source convolved with the {\sl Chandra} point spread function plus a 
double $\beta$~model \citep[\textit{projb model}]{Cro08}.
\citealt{Cro08} found no evidence for a temperature gradient in the hot gas.
The group is characterized by high temperature and low luminosity \citep{Fino06},
which taken together provide a very high level of entropy.
This might be a further sign of a large degree of impact of the AGN on the environment.

\subsection{3C\,353}
\label{353}

3C\,353 is an extended FR\,II radio source 
identified with a D-galaxy  embedded at the periphery of a
cluster of galaxies.
The best estimate for the inclination of the jets is $\approx$90\degrees \citep{Swain98}.
The eastern jet is slightly brighter and ends in a well-defined hot spot.
The radio lobes have markedly different morphologies: the eastern lobe 
is round with sharp edges, while the western lobe is elongated with an irregular shape.
The location of the source within the cluster is of particular interest for this work
and might account for the different shapes of the lobes. Fig.\,\ref{X}(d) shows
the XMM-Newton image overlaid on the radio contours. The image shows
only the NW part of the cluster, but it is clear that the radio source lies
on the edge of the X-ray emitting gas distribution.
so that the round eastern lobe is encountering a higher external density and is
probably also behind a larger column of Faraday-rotating material  
(\citealt{Iwa00}, \citealt{Good08}).  In particular,
the image published by \citet{Good08} shows that the gas density gradient
persists on larger scales.

\subsection{M\,84}
\label{84}

M\,84 is a giant elliptical galaxy 
located in the Virgo Cluster at about 400\,kpc from the core.
Optical emission-line imaging shows a disk of ionized gas around the nucleus,
with a maximum detected extent of $20 \times 7$\,arcsec$^2$ \citep*{Hansen85,Baum88,Bower97,Bower00}.
The radio emission of M\,84 (3C\,272.1) has 
an angular extension of about 3\,arcmin ($\simeq$~11\,kpc) and
shows an unresolved core in the nucleus of the galaxy, two resolved jets and a
pair of wide lobes \citep{LB87,laing11}. 
The inclination to the line-of-sight of the inner jet axis is 
$\sim$60\degrees, with the northern jet approaching, but
there is a noticeable bend in the counter-jet very close to the nucleus, which
complicates modelling \citep{LB11b}. After this bend, the jets remain straight
for $\approx$40\,arcsec, 
then both of them bend eastwards by $\sim$90\degrees and fade into 
the radio emission of the lobes. 

The morphology of the X-ray emission has a H-shape made up of shells of compressed gas
surrounding cavities coincident with both the radio lobes \citep{Fino08,Fino06,Fino01}.
This  shape, together with the fact that the initial bending of the radio 
jets has the same direction 
and is quite  symmetrical, suggests a combination of  interaction
with the radio plasma and motion of the galaxy within the cluster
\citep{Fino01}.
The ratio between the X-ray surface brightness
of the shells of the compressed gas and their surroundings is $\approx$3 and is almost 
constant around the source. The shells are regions of enhanced pressure and
density and low entropy: the amplitude of the density enhancements (a factor of $\approx$3)
suggests that they are produced by weak shock waves (Mach number ${\mathcal M}
\sim 1.3$) driven by the expanding lobes
\citep{Fino06}.

\section{Analysis of RM and depolarization images}
\label{sec:teq}

For a fully-resolved foreground Faraday screen, the $\lambda^2$ relation of
Eq.\,\ref{equarm} holds exactly and there is no change of degree of
polarization, $p$, 
with wavelength. Even in the presence of a small gradient of RM across the beam,
$\lambda^2$ rotation is observed over a wide range of polarization angle.
In this  case, the emission
tends to depolarize with increasing wavelength, following the Burn law  \citep{Burn66}:
\begin{equation}
\label{equadp}
p(\lambda)=p(0)\exp(-k\lambda^4),
\end{equation}
where $p(0)$ is the intrinsic value of the degree of polarization and
$k$=2$\arrowvert\nabla{RM}\arrowvert^2 \sigma^2$, with ${\rm FWHM}=2\sigma
(2\ln 2)^{1/2}$ \citep{Burn66,Tribble91a,laing08}. \\
Since $k \propto\arrowvert\nabla{RM}\arrowvert^2 $, Eq.\,\ref{equadp}
clearly illustrates that higher RM gradients across the beam generate higher $k$
values and in turn higher depolarization.
 
Our observational analysis is based on the following procedure.  We first
produced RM and Burn law $k$ images at two different angular resolutions for
each source and searched for regions with high $k$ or correlated 
RM and $k$ values, which could indicate the presence of internal Faraday
rotation and/or strong RM gradients across the beam. In regions with low $k$ where the
variations of RM are plausibly isotropic and random, we then used the structure
function (defined in Eq.\,\ref{sfunction}) to derive the power spectrum of the
RM fluctuations.  Finally, to investigate the
depolarization in the areas of isotropic RM, and hence the magnetic field power
on small scales, we made numerical simulations of the Burn law $k$ using the
model power spectrum with different minimum scales and compared
the results with the data.

The structure function is defined by
\begin{equation}
S(r_\perp)=\rm{<[RM({\bf r}_\perp + {\bf r}_\perp^\prime)-RM({\bf r}_\perp^\prime)]^2>}  
\label{sfunction}
\end{equation}
(\citealt*{SCS}; \citealt{MS96})
where ${\bf r_{\perp}}$ and ${\bf r'_{\perp}}$ are vectors in the plane 
of the sky and $\langle\rangle$ is an average over ${\bf r'_{\perp}}$.

We assume RM power spectra of the form:
\begin{eqnarray}
\hat{C}(f_\perp) & = & C_0 f_\perp^{-q} \makebox{~~~$f_\perp \leq f_{\rm max}$}
     \nonumber \\
     & = & 0 \makebox{~~~~~~~~~~$f_\perp > f_{\rm max}$} \label{eq-cutoff-pl} \nonumber 
\end{eqnarray}
where $f_\perp$ is a scalar spatial frequency and fit the observed structure
function (including the effect of the observing beam) using the Hankel-transform
method described by \citet{laing08} to derive the amplitude, $C_0$
and the slope, $q$. To constrain the RM structure on scales 
smaller than the beamwidth, 
we estimated the minimum scale of the 
best fitted field power spectrum, $\Lambda_{\rm min} = 1/f_{\rm max}$,
which predicts a mean value of $k$ consistent
with the observed one.

In this paper, we are primarily interested
in estimating the RM power spectrum over limited areas, and we made no attempt
to determine the outer scale of fluctuations.

The use of the structure function together with the Burn law $k$ 
represents a powerful technique to investigate the RM  power spectrum over
a wide range of spatial scales \citep{laing08,Guidetti10}. The two quantities
are complementary, in that the structure function allows us to determine the
power spectrum of the fluctuations on scales larger than the beamwidth, while 
the Burn law $k$ constrains fluctuations of RM below the resolution limit.

\section{Rotation measure images}
\label{sec:RM}

\begin{table}
\caption{Frequencies, bandwidths and angular resolutions used in the RM and 
Burn law $k$ images discussed in Sects.\,\ref{sec:RM} and \ref{dp}, respectively.
}
\label{nu}
\centering
\begin{tabular} {@{}c c c c}  
\hline
source & $\nu$ & $\Delta\nu$ & beam \\
       & [MHz]   &  [MHz]     &  [arcsec]  \\
\hline
0206+35 & 1385.1 & 25  & 1.2 \\
        & 1464.9 & 25   &  \\
        & 4885.1 & 50  &  \\
3C\,353 & 1385.0 & 12.5 & 1.3  \\
        & 1665.0 & 12.5 &   \\
        & 4866.3 & 12.5 &   \\
        & 8439.9 & 12.5 &   \\
3C\,270 & 1365.0 & 25  & 1.65\\
        & 1412.0 & 12.5  & \\
        & 4860.1 & 100 & \\
        & 1365.0  & 25 & 5.0\\
        & 1412.0 & 12.5 & \\
        & 1646.0 & 25  & \\
        & 4860.1 & 100 & \\
M\,84   & 1413.0 & 25 & 1.65 \\
        & 4885.1 & 50 &  \\ 
        & 1385.1 & 50 & 4.5 \\
        & 1413.0 & 25 &  \\
        & 1464.9 & 50 &  \\
        & 4885.1 & 50 &  \\
\hline
\end{tabular}
\end{table}

The RM images and associated rms errors were produced
by weighted least-squares fitting 
to the observed polarization angles $\Psi(\lambda)$ 
as a function of $\lambda^2$ (Eq.\,\ref{pang}) at three or four frequencies 
(Table~\ref{nu}, see also
\citealt{laing11} and \citealt{LB11a})
 using the {\sc RM} task in the {\sc AIPS} package.

Each RM map was calculated only 
at pixels with rms polarization-angle uncertainties $<$10\degrees
at all frequencies.
We refer only to the lower-resolution RM and $k$ images for 3C\,270 and M\,84 (Table~\ref{nu}),
as they  show more of the 
faint, extended regions of these sources and are 
fully consistent with the 
higher-resolution versions. The RM image of M\,84 is consistent with that shown
by \citet{LB87}, but is derived from four-frequency data and has a higher
signal-to-noise ratio.

In the fainter
regions of 0206+35 (for which only three frequencies are available and the
signal-to-noise ratio is relatively low), the {\sc RM} task occasionally failed
to determine the  n$\pi$ ambiguities in position angle correctly.
In order to remove these anomalies, we first produced a lower-resolution,
but high signal-to-noise RM image
by convolving the 1.2\,arcsec RM map 
to a beamwidth of 5\,arcsec FWHM.
From this map we derived the polarization-angle rotations 
at each of the three frequencies and
subtracted them from the observed 1.2\,arcsec polarization angle maps
at the same frequency to derive the residuals at high resolution.
Then, we fit the residuals without allowing any n$\pi$ ambiguities and added the
resulting RM's to the values determined at low resolution. This procedure allowed us to
obtain an RM map of 0206+35 free of significant deviations from $\lambda^2$
rotation and fully consistent with the 1.2-arcsec measurements.

\begin{figure*}
\centering
\includegraphics[width=9.5cm]{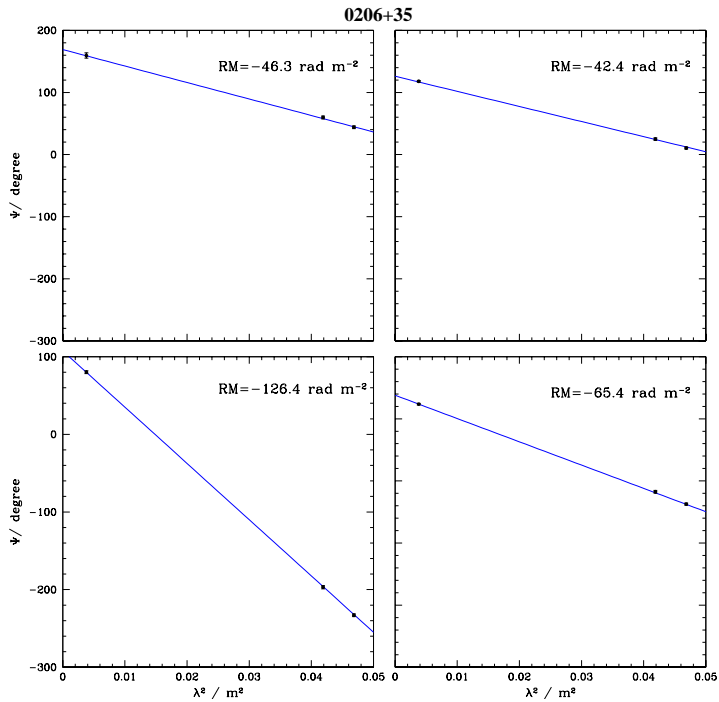}
\caption[]{Plots of ${\bf E}$-vector position angle $\Psi$ against $\lambda^2$
at representative points of the 1.2\,arcsec RM map of 0206+35. Fits to the relation 
$\Psi(\lambda) = \Psi_0 + {\rm RM}\lambda^2$ are shown. The values of
RM are given in the individual panels.}
\label{fittini}
\end{figure*}

We have verified that
the polarization angles accurately follow the relation
$\Delta \Psi \propto \lambda^2$ over the full range of position angle 
essentially everywhere except for small areas around the optically-thick cores: 
representative plots of $\Psi$ against $\lambda^2$ for 0206+35 
are shown in Fig.\,\ref{fittini}.
The lack of deviations from $\lambda^2$ rotation in all of the radio galaxies
is fully consistent with our assumption that the Faraday rotating medium is mostly external 
to the sources.

The RM maps are shown in Fig.\,\ref{RM}. The typical rms error on the fit is
$\approx$2\,rad\,m$^{-2}$. No correction for the Galactic 
contribution has been applied.
 
All of the RM maps show two-dimensional patterns, \textit{RM bands}, across the
lobes with characteristic widths ranging from 3 to 12\,kpc.  
Multiple bands parallel to each other are observed in the western lobe
of 0206+35, the eastern lobe of 3C\,353 and the southern lobe of M\,84.

In all cases, the iso-RM contours are straight and
perpendicular to the major axes of the lobes to a very good approximation: the
very straight and well-defined bands in the eastern lobes of both 0206+35
(Fig.\,\ref{RM}a) and 3C\,353 (Fig.\,\ref{RM}d) are particularly striking. The
entire area of M\,84 appears to be covered by a banded structure, while in the
central parts of 0206+35 and 3C\,270 and the western lobe of 3C\,353, regions of
isotropic and random RM fluctuations are also present.

We also derived profiles of \rmm along the radio axis of each source, averaging
over boxes a few beamwidths long (parallel to the axes), but extended
perpendicular to them to cover the entire width of the source. The boxes are all
large enough to contain many independent points.  The profiles are shown in
Fig.\,\ref{rmprof}. For each radio galaxy, we also plot an estimate of the Galactic contribution to the
RM derived from a weighted mean of the integrated RM's for non-cluster
radio sources within a surrounding area of 10\,deg$^2$ \citep{S-NKB}. In all cases, both positive and
negative fluctuations with respect to the Galactic value are present.

In 0206+35 (Fig.\,\ref{RM}a), the largest-amplitude bands are in the outer parts of the lobes, 
with a possible low-level band just to the NW of the core.
The most prominent band (with the most negative RM values) is in  the eastern
(receding) 
lobe, about 15 kpc from the core (Fig.\,\ref{rmprof}a). Its amplitude with
respect to the Galactic value is about 40\,rad\,m$^{-2}$.
This band must be associated with a strong ordered magnetic field component along the line of sight.
If corrected for the Galactic contribution, the two adjacent bands in the 
eastern lobe would have RM with opposite signs and the field
component along the line of sight must therefore reverse.

M\,84 (Fig.\,\ref{RM}b) displays an ordered RM pattern across the whole source, with two wide
bands of opposite sign having the highest absolute RM values.
There is also an abrupt change of sign across the radio core (see also
\citealt{LB87}).  The negative band in the northern lobe (associated with the approaching jet) has
a larger amplitude with respect to the Galactic value than the corresponding
(positive) feature in the southern lobe (Fig.\,\ref{rmprof}c). 

3C\,270 (Fig.\,\ref{RM}c) shows two large bands:  
one on the front end of the eastern lobe, the other in the middle of the western lobe.
The bands have opposite signs and contain the extreme positive and negative values of the observed RM.
The peak positive value is within the eastern band at the extreme end of the lobe
(Fig.\,\ref{rmprof}e).
 
The RM structure of 3C\,353  (Fig.\,\ref{RM}d) is highly asymmetric.
The eastern lobe shows a strong pattern, made up of four bands, with 
very straight iso-RM contours  which are almost exactly perpendicular 
to the source axis.  
As in 0206+35, 
adjacent bands have RM with opposite signs once corrected for the Galactic
contribution (Fig.\,\ref{rmprof}g).
In contrast, the RM distribution in the western lobe shows no sign of any banded
structure, and is consistent with random fluctuations superimposed on an almost
linear profile. It seems very likely that
the differences in RM morphology and axial ratio are both related to the external
density gradient (Fig.~\ref{X}d).

\begin{table*}
\caption{Properties of the RM bands: Col. 1 source name; Cols. 2\&3: overall \rmm and \srm;
Col. 4: Galactic \rmm; Col. 5: \rmm for each band; Col. 6: distance of the band midpoint 
from the radio core (positive distances are in the western direction for all sources, except
for M\,84, where they are in the northern direction); Col. 7: width of the band; Col. 8: maximum band amplitude.
}
\label{band}
\centering
\begin{tabular} {@{}l r r r r r r r r}  
\hline
source & \rmm & \srm & $\rm RM_{G}$ & band \rmm  & $\rm d_{c}$ & width & $A$\\
       &      &      &  [rad\,m$^{-2}$] &  [rad\,m$^{-2}$] & [kpc]    & [kpc] & [rad\,m$^{-2}$]  \\
\hline
0206+35 (4C\,35.03) & $-$77 & 23 & $-72$ & &    &  &   \\
        &       &    &     & $-$140 & -15 & 10 & 40  \\
        &       &    &     & $-$60  &-27 &  4 &  \\   
        &       &    &     &   34    &22  & 6  &    \\
        &       &    &     &   51    & 8  & 4  &    \\ 
3C\,353 & $-$56 & 24 & $-69$ &    &  & &     \\
        &       &    &     & 122  &-12  &  5 & 50 \\ 
        &       &    &     & 102 &-19  &  4    &       \\ 
        & & & & -40 & -23  &   4     &     \\ 
        & & & & 100 & -26  &    4     &      \\ 
3C\,270 & 14 & 10 & 12 & &  & & \\
        &    & &    & $-$8 & 20 & 12 & 10 \\ 
        & & & &32 & 37  &    11   &         \\ 
M\,84   & $-$2 & 15 & 2 & & & & \\
        &  & &   &$-$27 & 1   & 3 & 10     \\ 
        & & & &22 & -6 &   6 &           \\ 
\hline
\end{tabular}
\end{table*}

In Table \,\ref{band} the relevant geometrical features (size, distance from the radio core,
\rmm) for the RM bands are listed.

\section{Depolarization}
\label{dp}

In this section, we use ``depolarization'' in its conventional sense to mean
``decrease of degree of polarization with increasing wavelength'' and define DP
= $p_{\rm 1.4\,GHz}/p_{\rm 4.9\,GHz}$.
Using the Faraday code \citep{Murgia}, 
we produced images of Burn law $k$
by weighted least-squares fitting to $\ln p(\lambda)$
as a function of $\lambda^4$
(Eq.\,\ref{equadp}).
Only data with signal-to-noise ratio $>$4 in $P$
 at each frequency were included in the fits.
The Burn law $k$ images were produced
with the same angular resolutions as the RM images.
The 1.65\,arcsec
resolution
Burn law $k$ maps for M\,84 and
3C\,270 are consistent with the low-resolution ones, but add no additional
detail and are quite noisy. This could lead to significantly biased estimates for the
mean values of $k$ over large areas \citep{laing08}.
Therefore, as for the RM maps, we used only the Burn law $k$ images
at low resolution for these two sources.

The Burn law $k$ maps are shown in Fig.\,\ref{k}.
All of the sources show  low average values of $k$ (i.e.\ slight  
depolarization), suggesting little RM power on small 
scales. 
With the possible exception of the narrow filaments
of high $k$ in the eastern lobe of 3C\,353 (which might result from partially
resolved RM gradients at the band edges), none of the
images show any obvious structure related to the RM bands. 
For each source, we have also compared the RM and Burn law $k$ values 
derived by averaging over many small boxes covering the emission, and we find no
correlation between them.

\begin{figure*}
\centering
\includegraphics[width=12.5cm]{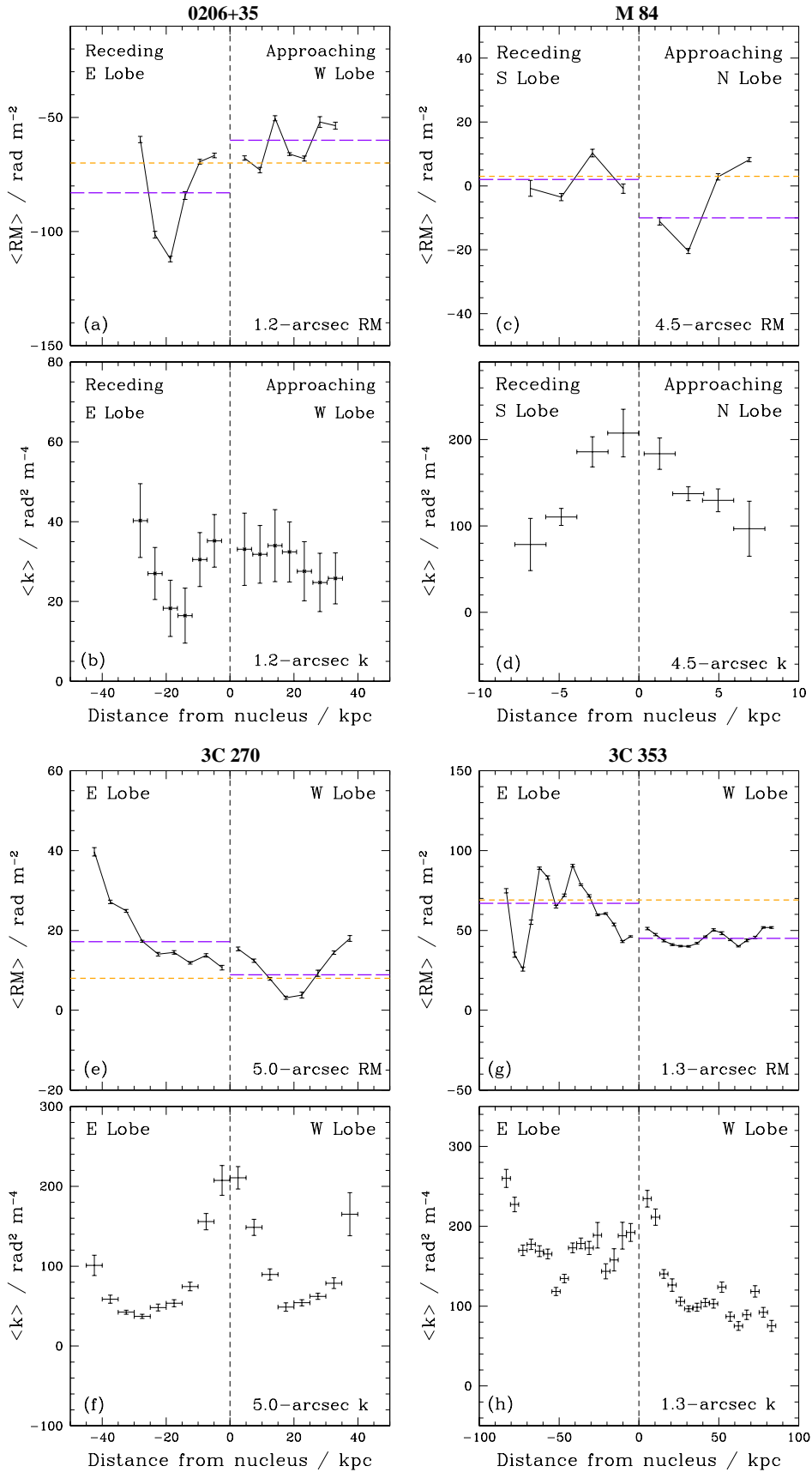}
\caption[]{Profiles of mean and rms RM and Burn law $k$ along the radio axis. 
The profiles have been derived by averaging over boxes
perpendicular to the radio axis of length
5\,kpc for all sources except  M\,84, for which the box length is 2\,kpc. 
The orange (dashed) and violet (long-dashed) lines represent the Galactic RM contribution and the
\rmm of the lobes, respectively.  
All unblanked pixels are included.
}
\label{rmprof}
\end{figure*}

We also derived profiles of $k$ (Fig.\,\ref{rmprof}b, d, f and h) 
with the same sets of boxes as for
the RM profiles in the same Figure.  These confirm that the values of $k$
measured in the centres of the RM bands are always low, but that there is little
evidence for any detailed correlation. 

The signal-to-noise ratio for 0206+35 is relatively low compared with that of
the other three sources, particularly at 4.9\,GHz (we need to use a small beam
to resolve the bands), and this is reflected in the high proportion of blanked
pixels on the $k$ image.  The most obvious feature of this image
(Fig.~\ref{k}a), an apparent difference in mean $k$ between the high-brightness
jets (less depolarized) and the surrounding emission, is likely to be an
artefact caused by our blanking strategy: points where the polarized signal is
low at 4.9\,GHz are blanked preferentially, so the remainder show artificially
high polarization at this frequency.  For the same reason, the apparent minimum
in $k$ at the centre of the deep, negative RM band (Fig.~\ref{rmprof}a and b) is
probably not significant. The averaged values of $k$ for 0206+35 are already
very low, however, and are likely to be slightly overestimated, so residual RM
fluctuations on scales below the 1.2-arcsec beamwidth must be very small.

\begin{figure*}
\centering
\includegraphics[width=15cm]{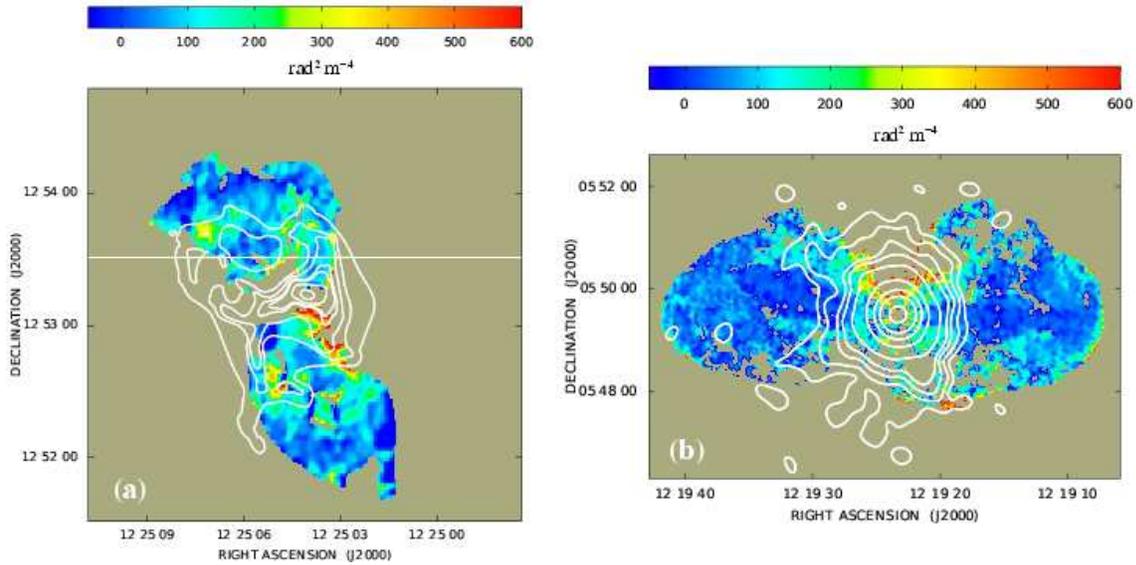}
\caption[]{Burn law $k$ images of M\,84 (a) and 3C\,270 (b)
overlaid on X-ray contours derived from {\sl Chandra} and XMM-Newton data, respectively.
}
\label{kx}
\end{figure*}

M\,84 shows one localised area of very strong depolarization
($k\sim$500\,rad$^2$\,m$^{-4}$, corresponding
to DP = 0.38)
at the base of the 
southern jet (Fig.~\ref{k}b).  There is no corresponding feature in the RM image
(Fig.~\ref{RM}b). The depolarization is likely to be associated with one of the 
shells of compressed
gas visible in the {\sl Chandra} image (Fig.~\ref{X}b),
implying significant magnetization with inhomogeneous field and/or density structure on
scales much smaller than the beamwidth, apparently independent of the
larger-scale field responsible for the RM bands.
This picture is supported by the good spatial coincidence
of the high $k$ region with a shell of compressed gas,
as illustrated in the overlay of the 4.5\,arcsec Burn law $k$ image
on the contours of the {\sl Chandra} data (Fig.\,\ref{kx}(a)). 
Cooler gas
associated with the emission-line disk might also be
responsible, but there is no evidence for spatial coincidence between 
enhanced depolarization and H$\alpha$ emission \citep{Hansen85}.
Despite the complex morphology of the X-ray emission 
around M\,84, its $k$ profile is very symmetrical,
with the highest values at the centre (Fig.\,\ref{rmprof}(d)).

3C\,270 also shows areas of very strong depolarization ($k\sim$550\,rad$^2$\,m$^{-4}$, corresponding to DP = 0.35)
close to the core and surrounding the inner and northern parts of both the radio lobes.
As for M\,84, the areas of high $k$  
are coincident with ridges in the X-ray emission which form the
boundaries of the cavity surrounding the lobes (Fig.\,\ref{kx}(b)).
The inner parts of this X-ray structure are described in more detail by
\citet{Worr10}, whose recent high-resolution {\sl Chandra} image clearly reveals ``wedges'' of low brightness
surrounding the radio jets.
As in M\,84, the most likely explanation is that a shell of denser gas
immediately surrounding the radio lobes is magnetized, with significant 
fluctuations of field strength and density on scales smaller than our 5-arcsec
beam, uncorrelated with the RM bands.   
The $k$ profile of 3C\,270 (Fig.\,\ref{rmprof}(f)) is very symmetrical,
suggesting that the magnetic-field and density distributions are also
symmetrical and consistent with an orientation close to the plane of the sky.
The largest values of $k$ are observed in the centre, coincident with the
features noted earlier and with the bulk of the X-ray emission (the high $k$ values
in the two outermost bins have low signal-to-noise and are not significant).

In the Burn law $k$ image of 3C\,353,
there is evidence for a straight and knotty region of high depolarization
$\approx$20\,kpc long and extending westwards from the core.
This region does not appear to be related 
either to the jets or to any other radio feature.
As in M\,84 and 3C\,270, the RM appears quite smooth over the area showing high
depolarization, again suggesting that there are two scales of structure, one much smaller than the beam, but
producing zero mean RM and the other very well resolved. 
In 3C\,353,  there is as yet no evidence for hot or cool
ionized gas associated with the enhanced depolarization 
(contamination from the very bright nuclear X-ray emission affects an 
area of 1\,arcmin radius around the core; \citealt{Iwa00}, \citealt{Good08}).

The $k$ profile of 3C\,353 (Fig.\,\ref{rmprof}(h)) shows a marked asymmetry,
with much higher values in the East. This is in the same sense as the difference
of RM fluctuation amplitudes (Fig.\,\ref{rmprof}(g)) and is also consistent with
the eastern lobe being embedded in higher-density gas.  The relatively high
values of $k$ within 20\,kpc of the nucleus in the Western lobe are due primarily to the
discrete region identified earlier.

\section{Rotation measure structure functions}
\label{sec:sfunc}

We calculated RM structure function for discrete regions of the sources 
where the RM fluctuations appear to be isotropic and random and for which we
expect the spatial variations of foreground thermal gas density, rms magnetic
field strength and path length to be reasonably small.  These are: the inner
26\,arcsec of the receding (Eastern) lobe of 0206+35, the inner 100\,arcsec of
3C\,270 and the inner 40\,arcsec of the western lobe of 3C\,353.  The selected
areas of 0206+35 and 3C\,270 are both within the core radii of the larger-scale
beta models that describe the group-scale X-ray emission and the
galaxy-scale components are too small to affect the RM statistics significantly
(Table~\ref{propX}).  In 3C\,353, the selected area was chosen to be small
compared with the scale of X-ray variations seen in Fig.~\ref{X}(d).  In all
three cases, the foreground fluctuations should be fairly homogeneous.  There
are no suitable regions in M\,84, which is entirely covered by the banded RM
pattern.

\begin{figure*}
\centering
\includegraphics[width=12cm]{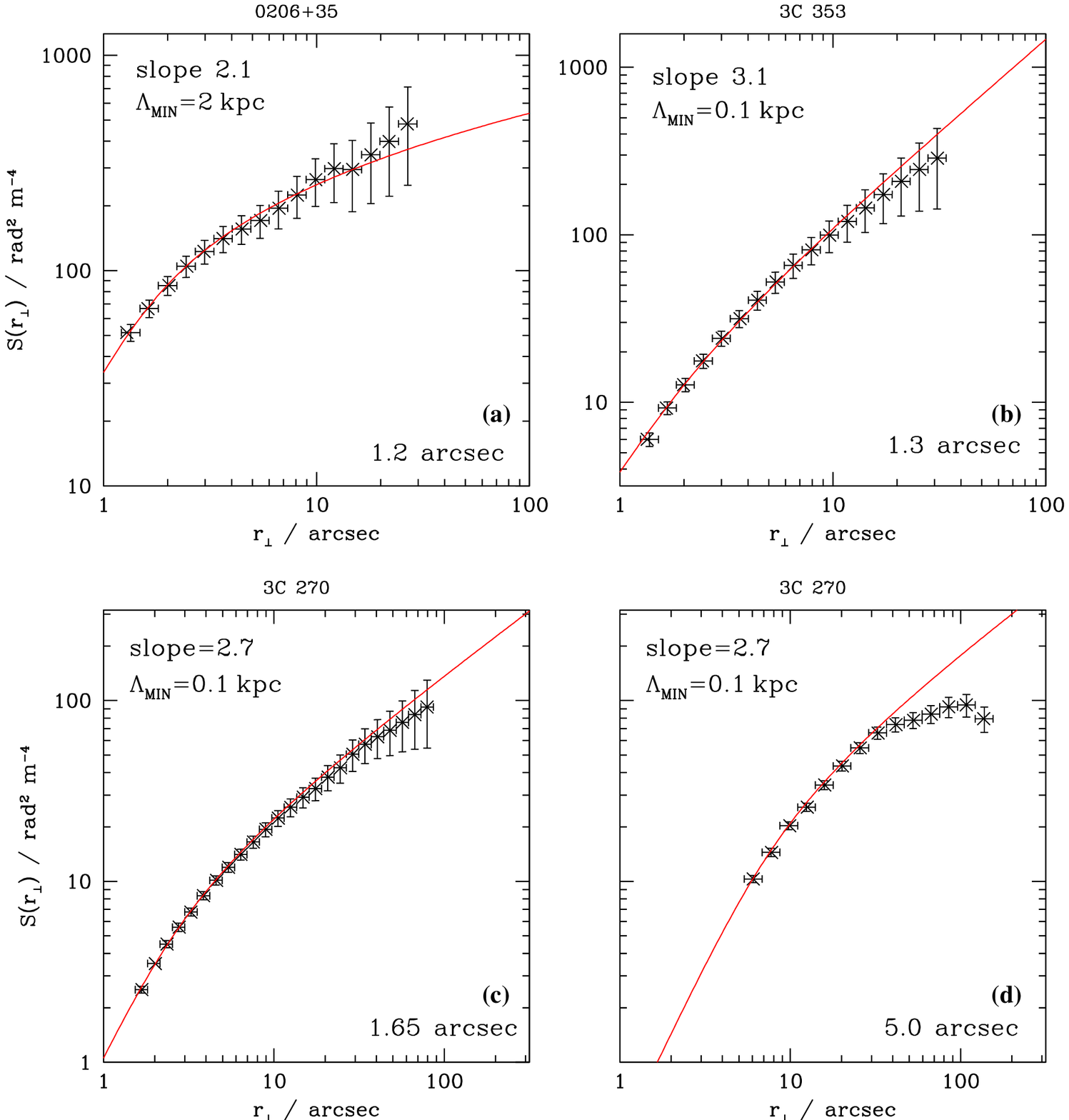}
\caption[]{Plots of the RM structure functions for the isotropic sub-regions
 of 0206+35,3C\,353 and 3C\,270 described in the text.
The horizontal bars represent the bin width and the crosses the centroids for 
data included in the bins. The red lines are the predictions for power law power 
spectra, including the effects of the convolving beam. 
The vertical error bars are the rms variations for the structure functions derived for
multiple realizations of the data.
The fitted structure functions for 3C\,270 are derived for the same power spectrum parameters.
}
\label{sfunc}
\end{figure*}

The structure functions, 
corrected for uncorrelated random noise by subtracting  2$\sigma_{\rm noise}^2$
\citep{SCS}, are shown in Fig.\,\ref{sfunc}.
All of the observed structure functions correspond to power spectra of approximately power-law
form over all or most of the range of spatial frequencies we sample. We initially assumed that
the power spectrum was described by Eq.\ref{eq-cutoff-pl} with no high-frequency
cut-off ($f_{\rm max} \rightarrow \infty$) and made 
least-squares fits to the structure functions, weighted by errors derived from
multiple realizations of the power spectrum on the observing grid, as described
in detail by \citet{laing08} and \citet{Guidetti10}.

The best-fitting slopes $q$ and amplitudes $C_0$ are
given in Table\,\ref{spectrum}.
All of the fitted power spectra are quite flat and have low amplitudes,
implying that there is little power in the isotropic and random component of rotation measure.
Indeed, the amplitudes of the largest-scale RM fluctuations 
sampled in this analysis is a few times less than that of the bands
(see Tables\,\ref{band} and \ref{spectrum}).
This suggests that the field responsible for the bands is  stronger as well as more
ordered than that responsible for the isotropic fluctuations.

\begin{table*} 
\caption{Power spectrum parameters for the individual sub-regions.
Col. 1: source name; Col. 2: angular resolution; Col. 3: slope $q$: Col. 4: amplitude $C_0$;
Col. 5: minimum scale; Col. 6: amplitude of the large scale
isotropic component; Col. 7: observed mean $k$; Col. 8: predicted mean $k$.
The power spectrum has not been computed for M\,84 (see Section~\ref{sec:sfunc}).
}
\label{spectrum}
\centering       
\begin{tabular}{@{}c c c c c c c c} 
\hline       
 Source     & FWHM & $q$ & $\log C_0$ & $\Lambda_{min}$ & $A_{iso}$  & $k^{obs}$ & $k^{syn}$ \\
               
            & [arcsec] &  &             & [kpc] & [rad\,m$^{-2}$] &  [rad$^2$\,m$^{-4}$] & [rad$^2$\,m$^{-4}$] \\
\hline      
0206+35 & 1.2   & 2.1 & 0.77 & 2   &  10   & 37 & 40 \\ 
3C\,270  &  1.65 & 2.7 & 0.90 & 0.3 & 5 & 30 & 26 \\
         &  5.0  & 2.7  & 0.90 & 0.3  & 5  & 71 & 64 \\
3C\,353  & 1.3   & 3.1  & 0.99   & 0.1 & 10  & 38 & 33   \\
M\,84    & 1.65  &      &        & $<$0.1 & & 25   &  \\
         & 4.5  &      &     &  $<$0.1    &  &  43  &    \\
\hline
\end{tabular}
\end{table*}

The structure functions for 0206+35 and 3C\,353 rise monotonically, indicating
that the outer scale for the random fluctuations must be larger than the maximum
separations we sample.  For 3C\,270, the structure function levels out at
$r_\perp \approx$ 100\,arcsec (15\,kpc; Fig.~\ref{sfunc}d).  This could be the
outer scale of the field fluctuations, but a better understanding of the geometry and external density
distribution would be needed before we could rule out the effects of large-scale
variations in path length or field strength (cf.\ \citealt{Guidetti10}).

In order to constrain RM
structure on spatial scales below the beamwidth, we estimated the depolarization
as described in Section~\ref{sec:teq}.
The fitted $k$ values are listed in Table\,\ref{spectrum}.
We stress that these values refer only to areas with isotropic fluctuations,
and cannot usefully be compared with the integrated depolarizations quoted in in Fig.\,\ref{RM}.

For M\,84, using the Burn law $k$ analysis and assuming that variation of Faraday rotation
across the 1.65-arcsec beam  
causes the residual depolarization, we find that $\Lambda_{\rm min} \la
0.1$\,kpc for any reasonable RM power spectrum.

\section{Rotation-measure bands from compression}
\label{model}

It is clear from the fact that the observed RM bands are perpendicular to the lobe axes
that they must be associated with an interaction between an expanding radio source
and the gas immediately surrounding it. One inevitable mechanism is
enhancement of field and density by the shock or compression wave surrounding
the source.\footnote{An alternative mechanism is the generation of non-linear
  surface waves \citep*{BCG90}. It is unlikely that this can
  produce large-scale bands, for the reasons given in Section~\ref{depol-mix}.} 
The implication of the presence of cavities in the X-ray gas distribution
coincident with the radio lobes is
that the sources are interacting strongly with the thermal gas, 
displacing rather than mixing with it (see \citealt{McN07} for a review).  For the sources in the present
paper, the X-ray observations of M\,84 \citep[Fig.\ref{X}b]{Fino08} and 3C\,270 \citep[Fig.\ref{X}c]{Cro08}
show cavities and arcs of enhanced brightness, corresponding to shells of
compressed gas bounded by weak shocks. The strength of any pre-existing field in the IGM, which will be
frozen into the gas, will also be enhanced in the shells. We therefore expect a significant
enhancement in RM.  A more extreme example of this effect will occur if the
expansion of the radio source is highly supersonic, in which case there will be a
strong bow-shock ahead of the lobe, behind which both the density and the field become 
much higher. Regardless of the strength of the shock, the field is
modified 
so that only the component in the plane of the shock 
is amplified and the post-shock field tends become
ordered  parallel to the shock surface.

The evidence so far suggests that shocks around radio sources of both  FR
classes are generally weak (e.g.\ \citealt{Forman05}, \citealt{Wilson06},
\citealt{Nulsen05}). There are only two examples in which highly supersonic
expansion has been inferred: the southern lobe of Centaurus\,A ($\mathcal{M}
\approx 8$; \citealt{Kraft03}) and NGC\,3801 ($\mathcal{M}
\approx 4$; \citealt{Cro07}).
There is no evidence that the sources described in  the present paper are
significantly overpressured compared with the surrounding IGM (indeed, the
synchrotron minimum pressure is systematically lower than the thermal pressure
of the IGM; Table \,\ref{propX}). The sideways expansion of the lobes is
therefore unlikely to be highly supersonic.  The shock Mach number estimated for all the
sources from ram pressure balance in the forward direction is also $\approx$~1.3.  This
estimate is consistent with that for M\,84 made by \citet{Fino06} and also with
the lack of detection of strong shocks in the X-ray data for the other sources.

In this section, we investigate how the RM
could be affected by compression. 
We consider a deliberately oversimplified picture in which the radio source expands into an IGM
with an initially uniform magnetic field, $B$.  This is the most favourable
situation for the generation of large-scale, anisotropic RM structures:  
in reality, the pre-existing field is likely to be highly disordered, or even isotropic,
because of turbulence in the thermal gas.  {\em We stress that we have not tried to
generate a self-consistent model for the magnetic field and thermal density, but
rather to illustrate the generic effects of compression on the RM structure}.

\begin{figure}
\centering
\includegraphics[width=6cm]{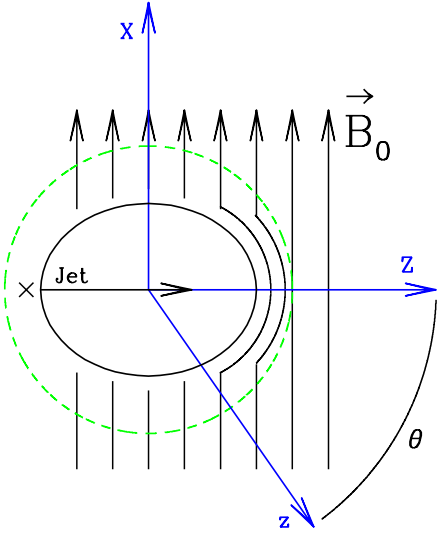}
\caption[]{Amplification and sweeping up of the magnetic field lines inside the compression region
defined by the projected circular green and dashed line. The driving expansion is along the $Z$-axis and
the $z$-axis represents a generic line-of-sight.
}
\label{prova}
\end{figure}

In this model the radio lobe is an ellipsoid with its major axis along the jet and is surrounded by a spherical shell 
of compressed material.
This shell is
centred at the mid-point of the lobe (Fig.\,\ref{prova})
and has a stand-off distance equal to 1/3 of the lobe semi-major axis at the
leading edge (the radius of the spherical compression is therefore equal to 4/3 of the lobe 
semi-major axis).
In the compressed region, the thermal density and the magnetic
field component in the plane of the spherical compression 
are amplified by the same factor, because of flux-freezing.
We use a coordinate system $xyz$
centred at the lobe mid-point, with 
the $z$-axis along the line of sight, so 
$x$ and $y$ are in the plane of the sky. The radial unit vector is $\bf{\hat{r}}$=$(\hat{x},\hat{y},\hat{z})$.
$\bf{B}$=$(B_{x},B_{y},B_{z})$
and $\bf{B}'$=$(B_{x}',B_{y}',B_{z}')$ are respectively the
pre- and post-shock magnetic-field vectors.
Then, we consider a coordinate system $XYZ$ still centred at 
the lobe mid-point, but rotated with respect to the $xyz$ system 
by the angle $\theta$ about the $y$ ($Y$) axis, so that
$Z$ is aligned with the major axis of the lobe. 
With this choice, $\theta$ 
is the inclination of the source with respect to the line-of-sight
(Fig.\,\ref{prova}).

After a spherical compression, 
the thermal density and field satisfy the equations:
\begin{eqnarray}
n_{\rm e}'~&=&\gamma n_{\rm e}\nonumber \\ 
 \bf{B'_{\perp}}~&=&\bf{B_{\perp}}=(\bf{B}\cdot\bf{r})\bf{\hat{r}}\nonumber\\
\bf{B'_{\parallel}}~&=&\gamma\bf{B_{\parallel}}=\gamma[\bf{B}-(\bf{B} \cdot\bf{r})\bf{\hat{r}}]
\label{freez}
\end{eqnarray}
where 
$n_{\rm e}$, $\bf{B_{\perp}}$ and $\bf{B_{\parallel}}$ represent
the initial thermal density and the components of the field perpendicular and parallel 
to the compression surface.
The same symbols with primes stand for post-shock quantities and $\gamma$ is the
compression factor.
The total compressed magnetic field is:
\begin{equation}
\bf{B'}=\bf{B'_{\parallel}}+\bf{B'_{\perp}}=\gamma\bf{B}+(1-\gamma)(\bf{B} \cdot\bf{r})\bf{\hat{r}}
\label{btot}
\end{equation}
The field strength after compression depends on the the angle between the
compression surface and the initial field. Maximum amplification occurs for a
field which is parallel to the surface, whereas a perpendicular field remains unchanged.
The post-shock field component along the line-of-sight becomes:
\begin{equation}
B'_{z}=\bf{B'}\cdot\hat{z}=\gamma\bf{B}\cdot\hat{z}+(1-\gamma)(\bf{B} \cdot \bf{r})(\bf{r}\cdot\hat{z}) 
\label{bz}
\end{equation}

We assumed that the compression factor, $\gamma=\gamma(Z)$, is a function of
distance $Z$ along the source axis from the centre of the radio lobe,
decreasing monotonically from a maximum value $\gamma_{\rm max}$ at the leading
edge to a constant value from the centre of the lobe as far as the core.  We
investigated values of $\gamma_{\rm max}$ in the range 1.5 -- 4 ($\gamma =4$
corresponds to the asymptotic value for a strong shock).  Given that there is no
evidence for strong shocks in the X-ray data for any of our sources, we have
typically assumed that the compression factor is $\gamma_{\rm max}=3$ at the
front end of the lobe, decreasing to 1.2 at the lobe mid-point and thereafter
remaining constant as far as the core.  A maximum compression factor of 3 is
consistent with the transonic Mach numbers $\mathcal{M} \simeq 1.3$ estimated
from ram-pressure balance for all of the sources and this choice is also
motivated by the X-ray data of M\,84, from which there is evidence for a
compression ratio $\approx 3$ between the shells and their surroundings
(Section\,\ref{84}).

We produced synthetic RM images
for different combinations of source inclination and direction of the pre-existing uniform field,
by integrating the expression
\begin{equation}
{\rm RM^{syn}}=\int_{lobe}^{R0}n'_{\rm e}B'_{z}dz.
\label{eq:comp}
\end{equation}
numerically. We assumed a constant value of 
$n_{\rm e}=2.4\times 10^{-3}$\,cm$^{-3}$ for the density of the pre-shock
material (the
central value for the group gas associated with 0206+35; Table~\ref{propX}), a lobe
semi-major axis of 21\,kpc (also appropriate for 0206+35) and an initial field
strength of 1\,$\mu$G.
The integration  limits were defined by the surface of the radio lobe and the 
compression surface. This is equivalent to assuming that there is no thermal gas within the radio lobe,
consistent with the picture suggested by  our inference of foreground Faraday
rotation and the existence of X-ray cavities and that Faraday rotation from
uncompressed gas is negligible.

\begin{figure*}

\centering
\includegraphics[width=16cm]{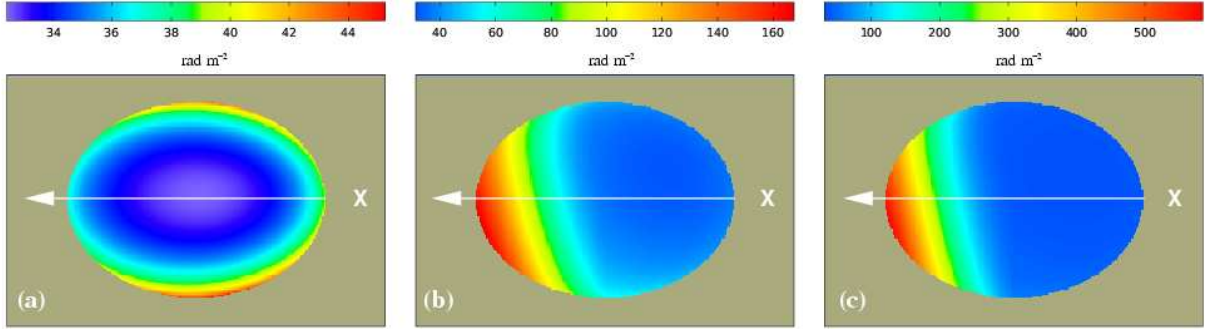}
\caption[]{Panel (a): synthetic RM for a receding lobe inclined by 
40\degrees to the line-of-sight and embedded in a uniform field inclined by
60\degrees  to the line-of-sight and by and 30\degrees
to the lobe axis on the plane of the sky.
Panels (b) and (c): as (a) but with a weak compression ($\gamma_{\rm max}=1.5$) and a
strong compression ($\gamma_{\rm max}=4$) of the density and field.
The crosses and the arrows indicate the position of the core and the lobe advance direction, 
respectively.
\label{ICM}
}
\end{figure*}

As an example, Fig.\,\ref{ICM} shows the effects of compression on the RM for the receding 
lobe of a source inclined by 40\degrees to the line of sight.  The initial field is pointing towards 
us with an inclination of 60\degrees with respect to the line-of-sight; its
projection on the plane of the sky makes an angle of 30\degrees with the $x$-axis.
Fig.\,\ref{ICM}(a) displays the RM produced without compression ($\rm
\gamma(Z)=1$ everywhere):
the RM structures are due only to 
differences in path length across the lobe.
Fig.\,\ref{ICM}(b) shows the consequence of adding a modest compression of
$\gamma_{\rm max} = 1.5$: structures similar to bands
are generated at the front end of the lobe and the range of the RM values is increased.
Fig.\,\ref{ICM}(c) illustrates the RM produced in case of the strongest possible
compression,
 $\rm \gamma_{\rm max} = 4$:
the RM structure is essentially the same as in Fig.\,\ref{ICM}(b), with a much 
larger range.

\begin{figure*}
\centering
\includegraphics[width=16cm]{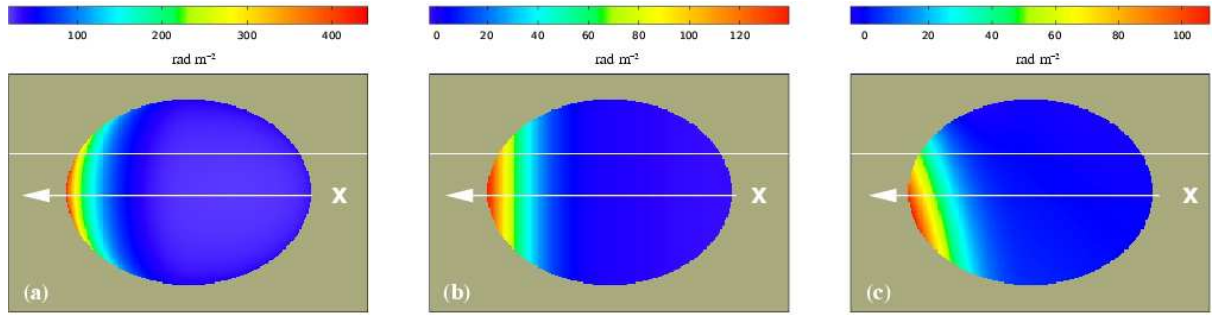}
\caption[]{Synthetic RM images for a lobe in the plane of the sky with
an ambient field which is uniform before compression. 
In panels (a) and (b), the field is along  the  
axis of the lobe in projection and inclined to
the line-of-sight by (a) 45\degrees and  (b) 90\degrees. In panel (c),  
the field is in the plane of the sky and misaligned by  45\degrees 
with respect to the lobe axis. The crosses and the arrows represent the radio core position and the 
lobe advance direction, respectively.
}
\label{90}
\end{figure*}

This very simple example shows that RM bands with amplitudes consistent with
those observed can plausibly be produced even by weak shocks in the IGM, but the iso-RM contours are neither
straight, nor orthogonal to the lobe axis and there are no reversals.  These constraints require specific
initial conditions, as illustrated in Fig.\,\ref{90}, where we show the RM for
a lobe in the plane of the 
sky. We considered three initial field configurations: 
along the line-of-sight (Fig.\,\ref{90}a), 
in the plane of the sky and parallel to the lobe axis (Fig.\,\ref{90}b) and 
in the plane of the sky, but inclined by 45\degrees to the lobe axis (Fig.\,\ref{90}c). 
The case closest to reproducing the observations is that displayed in Fig.\,\ref{90}(b), in which reversals 
and well defined and straight bands
perpendicular to the jet axis are produced for both of the lobes. In
Fig.\,\ref{90}(a), 
the structures are curved, while in  Fig.\,\ref{90}(c) 
the bands are perpendicular to the initial field direction, and therefore
inclined with respect to the lobe axis.

\begin{figure*}
\centering
\includegraphics[width=10.5cm]{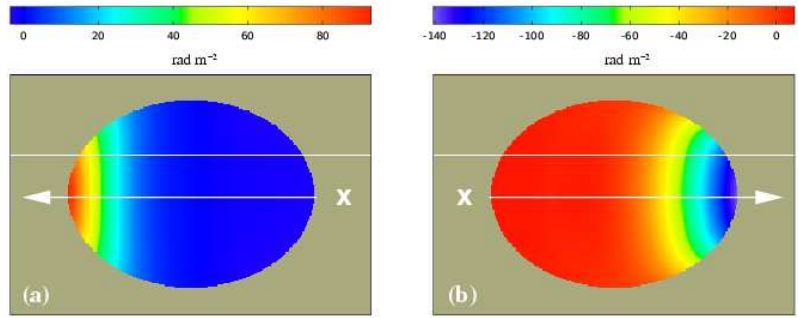}
\caption[]{Synthetic RM images of the receding (a) and approaching lobe (b) of a source 
inclined by 40\degrees with respect to the line-of-sight, in the case of a uniform ambient field.
The field is aligned with the radio jets in projection 
and inclined with respect to the line-of-sight by
90\degrees. The crosses and the arrows represent the radio core position and
the lobe advance direction, respectively.
}
\label{40}
\end{figure*}

For a source inclined by 40\degrees to the line of sight, we found 
structures similar to the observed bands only with an
initial field in the plane of the sky and parallel to the axis in projection
(Figs\,\ref{40}a and b; 
note that the synthetic RM images in this example have been made for each lobe separately,
neglecting superposition).

We can summarize the results of the spherical pure compression model as follows.
\begin{enumerate}
\item
An initial field with a component along the line-of-sight does not generate straight bands.
\item
The bands are orthogonal to the direction of the initial field projected on the
plane of the sky, so bands perpendicular to the lobe axis 
are only obtained with an initial field  aligned with the radio jet in projection.
\item
The path length (determined by the precise shape of the radio lobes) has a
second-order effect on the RM distribution (compare Figs.\,\ref{ICM}a and b).
\end{enumerate}

Thus, a simple compression model can generate bands with amplitudes similar to those observed
but reproducing their geometry requires implausibly special initial conditions, as we discuss
in the next section.

\section{Rotation-measure bands from a draped magnetic field}
\label{drap}

\subsection{General considerations}
\label{drap-qual}

That the pre-existing field is uniform, close to the plane of the sky and
aligned with the source axis in projection is implausible for obvious reasons:
\begin{enumerate}
\item
the pre-existing field cannot know anything about either the 
radio-source geometry or our line-of-sight and 
\item
observations of Faraday rotation in other sources and the theoretical inference
of turbulence in the IGM both require disordered initial fields.
\end{enumerate}
This suggests that the magnetic field must be aligned \textit{by}
rather than \textit{with} the expansion of the radio source. Indeed, the field
configurations 
which generate straight bands look qualitatively 
like the ``draping'' model proposed by \citet{DP08}, 
for some angles to the line-of-sight. The analysis of
Sections\,\ref{RM},~\ref{dp} and \ref{sec:sfunc} suggests that the magnetic
fields causing the RM bands are well-ordered,  
consistent with a stretching of the initial field that has erased much of the
small-scale structure while amplifying the large-scale component.  We next
attempt to constrain the geometry of the resulting ``draped'' field.

\subsection{Axisymmetric draped magnetic fields}
\label{3d}

A proper calculation of the RM from a draped magnetic-field configuration \citep{DP08} is outside
the scope of this paper, but we can start to understand the field geometry using
some simple approximations in which the field
lines are stretched along the source axis. We assumed initially that the field
around the lobe is axisymmetric, with components along and radially outwards
from the source axis, so that the RM pattern is independent of rotation about
the axis. 
\textit{It is important to stress that such an axisymmetric field is not physical}, as it requires a monopole and unnatural reversals,
it is nevertheless a useful benchmark for features of the field
geometry that are needed to account for the observed RM
structure.
We first considered field lines which are parabolae with a common
vertex on the axis ahead of the lobe.
For field strength and density both decreasing
away from the vertex, we found that RM structures,
with iso-contours similar to arcs, rather than bands,  were generated 
only for the approaching lobe of an inclined source.
Such anisotropic RM structures were not produced
in the receding lobes, nor for sources
in the plane of the sky. Indeed, 
in order to generate any narrow,
transverse RM structures such as arcs or bands,
the line-of-sight must pass through a foreground region in which the 
field lines show significant curvature, which occurs
only for an approaching lobe in case of a parabolic field geometry.
This suggested that we should consider field lines which are families of
ellipses centred on the lobe, again with field strength and density decreasing
away from the leading edge.  This indeed produced RM structures in both lobes
for any inclination, but the iso-RM contours were arcs, not straight lines. 
Because of the non-physical nature of these axisymmetric field models,
the resulting RM images are deliberately not shown in this paper. 
In order to quantify the departures from straightness of the iso-RM
contours, we measured the ratio of the predicted RM values at the centre and edge of 
the lobe at constant $Y$, at different distances along the source axis, $X$, for both
of the example axisymmetric field models. The ratio, which is 1 for
perfectly straight bands, varies from 2 to 3 in both cases, depending on
distance from the nucleus. This happens because the variations in
line-of-sight field strength and density do not compensate accurately for
changes in path length.  We believe that this problem is generic to any
axisymmetric field configuration. \\
The results of this section suggest that the field configuration required
to generate straight RM bands perpendicular to the projected lobe advance
direction has systematic curvature in the field lines (in order to produce
a modulation in RM) without a significant dependence on azimuthal angle
around the source axis).  We therefore investigated a structure in which
elliptical field lines are wrapped around the front of the lobe, but in a
two-dimensional rather than a three-dimensional configuration.

\subsection{A two-dimensional draped magnetic field}
\label{2d}

\begin{figure*}
\centering
\includegraphics[width=13cm]{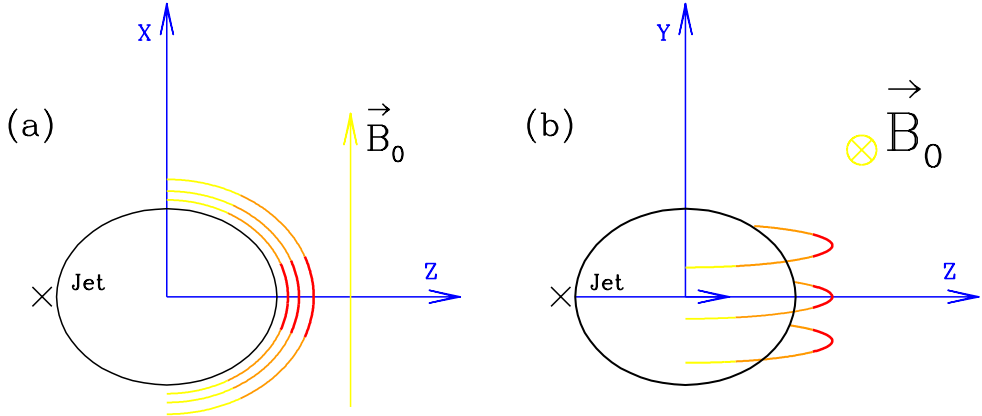}
\caption[]{Geometry of the two-dimensional draped model for the magnetic field,
described in Sec.\,\ref{2d}. Panel (a): seen in the plane of the direction of 
the ambient magnetic field.
Panel (b): seen in the plane perpendicular to the ambient field; only 
the innermost family of ellipses is plotted.
For clarity, field lines behind the lobe are omitted.
The colour of the curves indicates the strength of the field, decreasing from red to yellow. 
The crosses represent the radio core position. The coordinate system is the same as in Fig.~\ref{prova} and the line of sight is in the $X - Z$ plane.
}
\label{famab}
\end{figure*}

\begin{figure*}
\centering
\includegraphics[width=17cm]{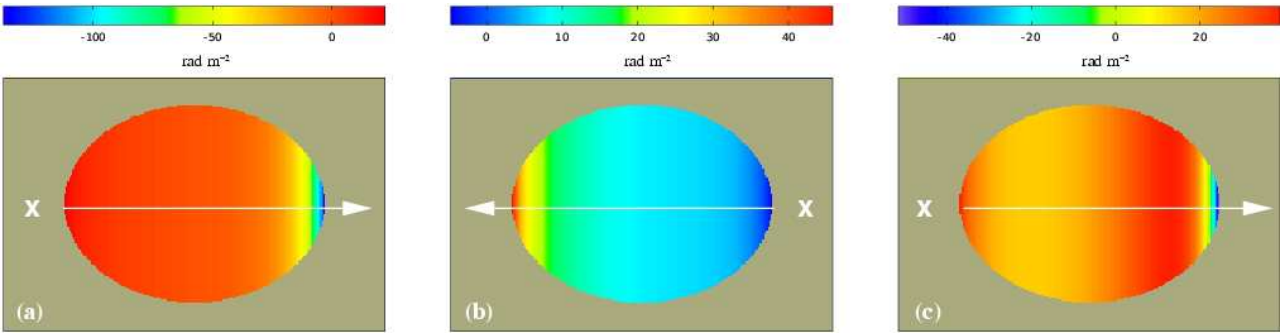}
\caption[]{Synthetic RM images produced for a two-dimensional draped model as
described in Section~\ref{2d}. The field lines are a family of ellipses draped
around the lobe as in Fig.~\ref{famab}. Panel (a): lobe inclined at 
90\degrees to the line-of-sight. (b) receding and (c) approaching lobes of a source
inclined by 40\degrees to the line-of-sight. The crosses and the arrows represent the radio core position 
and the lobe advance direction.}
\label{synfam}
\end{figure*}

We considered a field with a two-dimensional geometry, in which the
field lines are families of ellipses in planes of constant $Y$, as sketched in
Fig.\,\ref{famab}. The field structure is then independent of $Y$. 
The limits of integration  are given by the lobe surface and an
ellipse whose major axis is 4/3 of that of the lobe.
Three example RM images are shown in
Fig.~\ref{synfam}. The assumed density ($2.4 \times 10^{-3}$\,cm$^{-3}$) and
magnetic field (1\,$\mu$G) were the same as the pre-shock values for the
compression model of Section~\ref{model} and the lobe semi-major axis was again 21\,kpc.
Since the effect of path length is very small (Section~\ref{model}), the RM is also independent
of $Y$ to a good approximation. The combination of elliptical field lines and
invariance with $Y$ allows us to produce straight RM bands perpendicular to the
projected lobe axis for any source inclination. 
Furthermore, this model generates 
more significant reversals of the RM 
(e.g.\ Fig.\,\ref{synfam}c) than those obtainable with pure compression 
(e.g.\ Fig.\,\ref{40}b).

We conclude that a field model of this generic type 
represents the simplest way to produce RM bands with the observed
characteristics in a way that does not require improbable initial conditions. 
The invariance of the field with the $Y$ coordinate
is an essential point of this model, suggesting that
the physical process responsible for the draping and stretching
of the field lines must act on scales larger than the 
radio lobes in the $Y$ direction.
\subsection{RM reversals}
\label{helic}

\begin{figure}
\centering
\includegraphics[width=6cm]{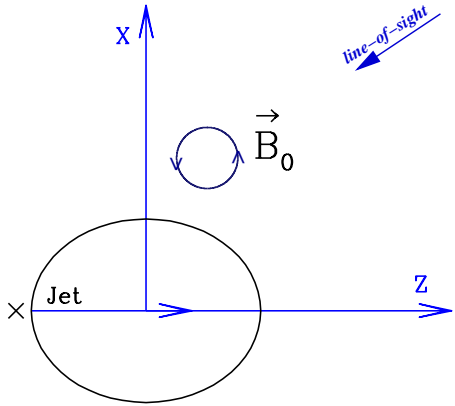}
\caption[]{Geometry of a two-dimensional toroidal magnetic field as described in
Section~\ref{2d}, seen in the plane normal to its axis. The cross represents the
radio core position.
}
\label{circle}
\end{figure}

The two-dimensional draped field illustrated in the previous section reproduces
the geometry of the observed RM bands very well, but can only generate a single
reversal, which must be very close to the front end of the approaching lobe, where the elliptical
field lines bend most rapidly. We observe a prominent reversal in the receding
lobe of 0206+35 (Fig.~\ref{RM}a) and multiple reversals across the eastern lobe
of 3C\,353 and in M\,84 (Fig.~\ref{RM}b and d).  The simplest
way to reproduce these is to assume that the draped field also has reversals,
presumably originating from a more complex initial field in the IGM.  One
realization of such a field configuration would be in the form of multiple
toroidal eddies with radii smaller than the lobe size, as sketched in
Fig.\,\ref{circle}.  Whatever the precise field geometry, we stress that the
straightness of the observed multiple bands again requires a two-dimensional
structure, with little dependence on $Y$.

\section{Discussion}
\label{discuss}

\subsection{Where do bands occur?}
\label{many}

The majority of published  RM images of radio galaxies 
do not show bands or other kind of anisotropic structure, but are characterized by 
isotropic and random RM 
distributions (e.g.\ \citealt{laing08,Guidetti10}).  
On the other hand, we have presented observations of  RM bands in four radio galaxies 
embedded in different environments and with a range of jet  
inclinations with respect to the line-of-sight.  Our sources are not drawn from
a complete sample, so any quantitative estimate of the incidence of
bands is premature, but we can draw some preliminary conclusions.

The simple two-dimensional draped-field model developed in Section~\ref{2d} only
generates RM bands when the line-of-sight intercepts the volume containing
elliptical field lines, which happens for a restricted range of rotation around
the source axis. At other orientations, the RM from this field configuration
will be small and the observed RM may well be dominated by material at larger
distances which has not been affected by the radio source.  We therefore expect a minority of sources with
this type of field structure to show
RM bands and the remainder to have weaker, and probably isotropic, RM
fluctuations. In contrast, the three-dimensional draped field
model proposed by \citet{DP08} predicts RM bands {\em parallel} to the source axis for
a significant range of viewing directions: these have not (yet) been observed.

The prominent RM bands described in the present paper occur only in {\em lobed}
radio galaxies. In contrast, well-observed radio sources with tails and plumes seem to be free of bands or
anisotropic RM structure (e.g. 3C\,31, 3C\,449; \citealt{laing08,Guidetti10}).
Furthermore, the lobes which show bands are all quite round and show evidence
for interaction with the surrounding IGM.  It is particularly striking that the bands in
3C\,353 occur only in its eastern, rounded, lobe.  The implication is that RM
bands occur when a lobe is being actively driven by a radio jet into a region of
high IGM density.  Plumes and tails, on the other hand, are likely to be rising
buoyantly in the group or cluster and we do not expect significant compression,
at least at large distances from the nucleus.

\subsection{RM bands in other sources}
\label{other-sources}

The fact that RM bands have so far been observed only in a few radio sources
may be a selection effect: much  RM analysis has been carried out for galaxy clusters,
in which most of the sources are tailed (e.g. \citealt{Blanton03}). With a few
exceptions like Cyg\,A (see below, Section~\ref{cyg}), lobed FR\,I and FR\,II
sources have not been studied in detail.

\subsubsection{Cygnus\,A}
\label{cyg}

Cygnus\,A is a source in which we might expect to observed RM bands, by analogy
with the sources discussed in the present paper:
it has wide and round lobes and {\sl Chandra} X-ray data have shown the presence of 
shock-heated gas and cavities \citep{Wilson06}. RM bands, roughly perpendicular
to the source axis, are indeed seen in both lobes \citep{DCP87,CT02}, but interpretation is
complicated by the larger random RM fluctuations and the strong
depolarization in the eastern lobe.  A semi-circular RM feature around one of
the hot-spots in the western lobe has been attributed to compression by the
bow-shock \citep*{CPD88}.

\subsubsection{Hydra A}
\label{hydra}

\citet{CT02} have claimed evidence for RM bands in the northern lobe of
Hydra\,A. 
The {\sl Chandra} image \citep{McN2000} shows a clear cavity with sharp edges 
coincident with the radio lobes and an absence of  shock-heated gas, just as in our sources.
Despite the classification as a tailed source, it may well be that there is
significant compression of the IGM. Note, however, that the RM image is not well
sampled close to the nucleus.

\subsubsection{3C\,465}
\label{465}
The RM image of the tailed source 3C\,465 published by \citet{EO02} 
shows some evidence for bands, but the colour scale was deliberately chosen to highlight the
difference between positive and negative values, thus making it difficult to see
the large gradients in RM expected at band edges. The original RM image (Eilek, private
communication) suggests that the band in the western tail of 3C\,465 is similar
to those we have identified.
It is plausible that magnetic-field draping happens in wide-angle tail sources
like 3C\,465 as a result of bulk motion of the IGM within the cluster potential
well, as required to bend the tails. It will be interesting to search for RM
bands in other sources of this type and to find out whether there is any
relation between the iso-RM contours and the flow direction of the IGM. 

\subsection{Foreground isotropic field fluctuations}
\label{iso}

The coexistence in the RM images of our sources of anisotropic patterns with areas
of isotropic fluctuations suggests that the Faraday-rotating medium 
has at least two components: one local to the source, where its motion 
significantly affects the surrounding medium, draping the field, and the other 
from material on group or cluster scales which has not felt the effects of the source. 
This raises the possibility that  turbulence in the foreground
Faraday rotating medium might ``wash out'' RM bands, thereby making them
impossible to detect.  The isotropic RM fluctuations observed across our sources are  
all described by quite flat power spectra
with low amplitude (Table\,\ref{spectrum}). The random small-scale structure of the field 
along the line-of-sight essentially averages out, and there is very little power
on scales comparable with the bands. 
If, on the other hand, the isotropic  field had a steeper power spectrum with significant
 power on scales similar to the bands, then its contribution might become dominant.

We first produced synthetic RM images for 0206+35 including a random component
derived from our best-fitting power spectrum (Table\,\ref{spectrum}) in order to
check that the bands remained visible.  We assumed a minimum scale of 2\,kpc
from the depolarization analysis for 0206+35 (Sec.\,\ref{dp}), and a maximum
scale of $\Lambda_{\rm max} = 40$\,kpc, consistent with the continuing rise of
the RM structure function at the largest sampled separations, which requires
$\Lambda_{\rm max} \ga 30$\,arcsec ($\simeq$20\,kpc).  The final synthetic RM is
given by:
\begin{equation}
{\rm RM^{\rm syn}}={\rm RM^{\rm drap}}+{\rm RM^{\rm icm}}=\int n'_{\rm e}B'_{z}dz+\int n_{\rm e}B_{z}dz
\label{eq:ICM}
\end{equation}
where
$\rm RM^{\rm drap}$ and $\rm RM^{\rm icm}$ 
are the RM due to the draped and isotropic fields, respectively. 
The terms $n'_{\rm e}$ and $B'_{\rm z}$ are respectively 
the density and field component along the line-of-sight in the draped region.
The integration limits of the term $\rm RM^{\rm drap}$
were defined by the surface of the lobe and the 
draped region,
while that of the term $\rm RM^{\rm icm}$ starts 
at the surface of the draped region and extends to 3 times the core radius of
the X-ray gas 
(Table\,\ref{propX}).
For the electron gas density $n_{\rm e}$ outside the draped region we assumed 
the beta-model profile of 0206+35
(Table\,\ref{propX}) and for the
field strength a radial variation of the form \citep[and references therein]{Guidetti10}.
\begin{equation}\label{br}
\langle B^2(r)\rangle^{1/2} = B_{0} \left[\frac{n_{\rm e}(r)}{n_0}\right]^{~\eta}
\end{equation}
where $B_{0}$ is the rms magnetic field strength at the group centre.
We took a draped magnetic field strength of 1.8\,$\mu$G, in order to match the 
amplitudes for the RM bands in both lobes of 0206+35, and assumed the same value
for $B_{0}$. 

\begin{figure*}
\centering
\includegraphics[width=12cm]{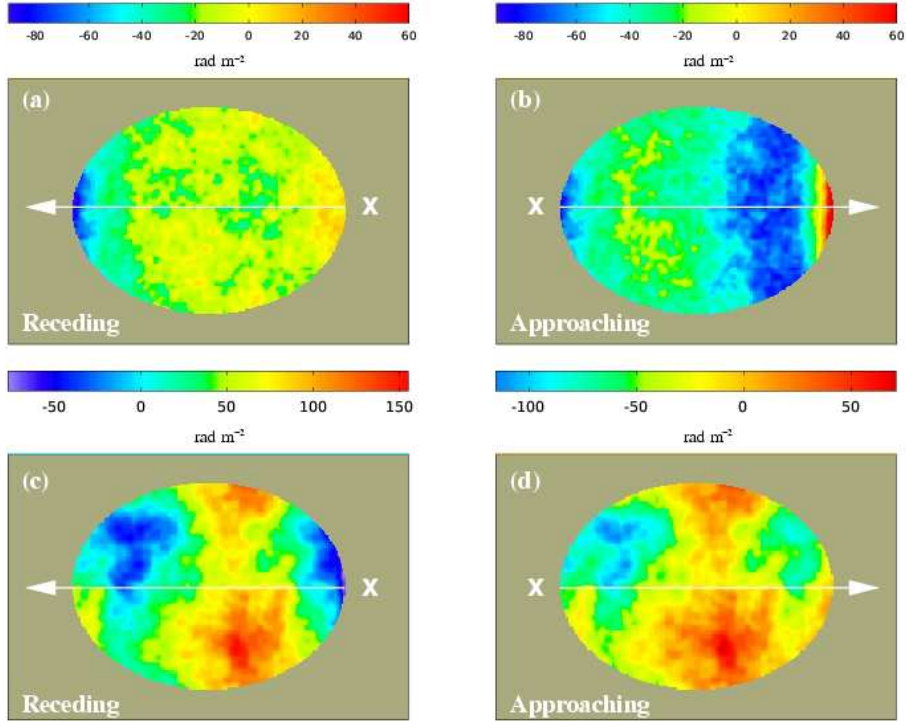}
\caption[]{Synthetic RM images of the lobes of 0206+35,
 produced by the sum of a draped field local to the source and an isotropic, random field
spread through the group volume with the best-fitting power spectrum found 
in Sec.\,\ref{sec:sfunc} (panels (a) and (b)) and 
with a Kolmogorov power spectrum (panels (c) and (d)). Both power
spectra have the same high and low-frequency cut offs and the central magnetic
field strengths are also identical. 
The crosses and the arrows represent the radio core position and 
the lobe advance direction.
}
\label{drapicm}
\end{figure*}

Example RM images, shown in Figs.\,\ref{drapicm}(a) and (b), should be
compared with those for the draped field alone (Figs.\,\ref{synfam}b and c,
scaled up by a factor of 1.8 to account for the difference in field strength) and
with the observations (Fig.\ref{RM}a) after correction for Galactic foreground
(Table\,\ref{band}). 
The model is self-consistent: the flat power spectrum found for 0206+35
does not give coherent RM structure which could interfere with the RM bands,
which are still visible.

We then replaced the isotropic component with one having a Kolmogorov power spectrum ($q=11/3$).
We assumed identical minimum and maximum scales (2 and 40\,kpc) as for
the previous power spectrum and took the same central field strength
($B_{0}$=1.8$\mu$G) and radial variation (Eq.\,\ref{br}). Example realizations
are shown in Fig.\,\ref{drapicm}(c) and (d).
The bands are essentially invisible in the presence of foreground RM fluctuations with a
steep power spectrum out to scales larger than their  widths.

It may therefore be that the RM bands in our sources are especially prominent
because the power spectra for the isotropic RM fluctuations have unusually low
amplitudes and flat slopes. We have established these parameters directly for
0206+35, 3C\,270 and 3C\,353; M\,84 is only 14\,kpc in size and is located far from the core of the Virgo
cluster, so it is plausible that the cluster contribution to its RM is small and constant.   

We conclude that the detection of RM bands
could be influenced by the relative amplitude and scale of the fluctuations of the isotropic
and random RM component compared with that from any draped field, and that
significant numbers of banded RM structures could be masked by isotropic
components with steep power spectra. 

\subsection{Asymmetries in RM bands}
\label{asy}

The well-established correlation between RM variance and/or depolarization and
jet sidedness observed in FR\,I and FR\,II radio sources is interpreted as an
orientation effect: the lobe containing the brighter jet is on the near side,
and is seen through less magnetoionic material (e.g.\ \citealt{L88,Morg97}).
For sources showing RM bands, it is interesting to ask whether the asymmetry is
due to the bands or just to the isotropic component.

In 0206+35, whose jets are inclined by $\approx$~40\degrees to the line-of-sight,
the large negative band on the receding side has the highest RM (Fig.\,\ref{rmprof}).
This might suggest that the RM asymmetry is due to the bands, and therefore to  the local draped field.
Unfortunately, 0206+35 is the only source displaying this kind of asymmetry.
The other ``inclined'' source, M\,84, ($\theta\approx$~60\degrees) 
does not show such asymmetry:  on the contrary the RM amplitude is quite symmetrical. 
In this case, however, the relation between the (well-constrained) inclination
of the inner jets,
and that of the lobes could be complicated: both jets bend by
$\approx$~90\degrees  at distances of about 50\,arcsec from the nucleus
\citep{laing11}, so that we cannot establish the real orientation of the lobes with respect to 
the plane of the sky. The low values of the jet/counter-jet ratios in 3C\,270 and 3C\,353 suggest that
their axes are close to the plane of the sky, so that little
orientation-dependent RM asymmetry would be expected.
Indeed, the lobes of 3C\,270 show similar RM amplitudes, while the large 
asymmetry of RM profile of 3C\,353 is almost certainly due to a higher column
density of thermal gas in front of the eastern lobe. 
Within our small sample,  there is therefore no convincing evidence for higher 
RM amplitudes in the bands on the receding side, but neither can such an effect be ruled out. 

In models in which an ordered field is draped around the radio lobes, the
magnitude of any RM asymmetry depends on the field geometry as well as the path
length (cf.\ \citealt{laing08} for the isotropic case).  For instance, the
case illustrated in Figs~\ref{synfam}(b) and (c) shows very little asymmetry
even for $\theta$=~40\,\degrees.  The presence of a systematic asymmetry in the
banded RM component could therefore be used to constrain the geometry. 

\subsection{Enhanced depolarization and mixing layers}
\label{depol-mix}

A different mechanism for the generation of RM fluctuations was
suggested by \citet{BCG90}.  They argued that large-scale nonlinear surface waves
could form on the surface of a radio lobe through the merging of smaller waves
generated by Kelvin-Helmholtz instabilities and showed that RM's of roughly the
observed magnitude would be produced if a uniform field inside the lobe was
advected into the mixing layer. This mechanism is unlikely to be able to
generate large-scale bands, however: the predicted iso-RM contours are only
straight over parts of the lobe which are locally flat, even in the unlikely
eventuality that a coherent surface wave extends around the entire lobe.  

The idea that a mixing layer generates high Faraday rotation may instead be
relevant to the anomalously high depolarizations associated with regions of
compressed gas around the inner radio lobes of M\,84 and 3C\,270
(Section~\ref{dp} and Fig.~\ref{kx}).  We have argued that the fields
responsible for the depolarization are tangled on small scales, since they
produce depolarization without any obvious effects on the large-scale Faraday
rotation pattern. It is unclear whether the level of turbulence within the
shells of compressed gas is sufficient to amplify and tangle a pre-existing
field in the IGM to the level that it can produce the observed depolarization; a
plausible alternative is that the field originates within the radio lobe and
mixes with the surrounding thermal gas.

\section{Conclusions and outstanding questions}
\label{concl}

In this work we have analysed and interpreted the Faraday rotation across the
lobed radio galaxies 0206+35, 3C\,270, 3C\,353 and M\,84, located in
environments ranging from a poor group to one of the richest clusters of
galaxies (the Virgo cluster).  The RM images have been produced at resolutions
ranging from 1.2 to 5.5\,arcsec FWHM  using Very Large Array data at
multiple frequencies.  All of the RM images show peculiar banded patterns across
the radio lobes, implying that the magnetic fields responsible for the Faraday
rotation are anisotropic.  The RM bands coexist and contrast with areas
of patchy and random fluctuations, whose power spectra have been estimated using
a structure-function technique.  We have also analysed the variation of degree
of polarization with wavelength and compared this with the predictions for the
best-fitting RM
power spectra in order to constrain the minimum scale of magnetic turbulence.
We have investigated the origin of the bands by making 
synthetic RM images using simple models of the interaction between radio
galaxies and the surrounding medium and have estimated the geometry and strength
of the magnetic field.

Our results can be summarized as follows.
\begin{enumerate}
\item The lack of deviation from $\lambda^2$ rotation over a wide range of
  polarization position angle and the lack of associated depolarization together
  suggest that a foreground Faraday screen 
  with no mixing of radio-emitting and
  thermal electrons is responsible for the observed RM in the bands and
elsewhere
  (Section~\ref{sec:RM}).
\item The dependence of the degree of polarization on wavelength
  is well fitted by a Burn law, which is also consistent with (mostly resolved)  pure foreground
  rotation (Section~\ref{dp}). 
\item The RM bands are typically 3 -- 10\,kpc wide and have amplitudes of 10 --
  50\,rad\,m$^{-2}$ (Table~\ref{band}). The maximum deviations of RM from the
  Galactic values are observed at the position of the bands.  Iso-RM contours
  are orthogonal to the axes of the lobes.  In several cases, neighbouring bands
  have opposite signs compared with the Galactic value and the line-of-sight
  field component must therefore reverse between them.
 \item  An analysis of the profiles of \rmm  and depolarization along the source axes
        suggests that there is very little small-scale RM structure within the bands.
\item
The lobes against which bands are seen have unusually small axial ratios (i.e.\
they appear round in projection; Fig.~\ref{RM}).  
In one source (3C\,353) the two lobes differ
significantly in axial ratio, and only the rounder one shows RM bands.  This lobe is
on the side of the source for which the external gas density is higher.
 \item Structure function and depolarization analyses show that
       flat power-law power spectra with low amplitudes and high-frequency cut-offs are
       characteristic of the areas which show isotropic and random RM
       fluctuations, but no bands 
       (Section~\ref{sec:sfunc}).
\item
There is evidence for source-environment interactions, such as large-scale
asymmetry (3C\,353) cavities and 
shells of swept-up and compressed material (M\,84, 3C\,270) in all three sources for which
high-resolution X-ray imaging is available.
\item Areas of strong depolarization are found around the edges of the radio
  lobes close to the nuclei of  3C\,270 and M\,84. These are probably associated
  with shells of compressed hot gas. The absence of large-scale changes in
  Faraday rotation in these features suggests that the field must be tangled on
  small scales (Section~\ref{dp}).
 \item The comparison of the amplitude of \rmm with that of the structure functions
       at the largest sampled separations is consistent with an amplification
       of the large scale magnetic field component at the position of the bands.       
\item  We produced synthetic RM images from radio lobes expanding into an ambient 
       medium containing thermal material and magnetic field, first considering
       a pure compression of both thermal density and field, and then including
       three- and two-dimensional stretching (``draping'') of the field lines along the direction
        of the radio jets (Sects.~\ref{model} and \ref{drap}).     
       Both of the mechanisms are able to generate anisotropic RM structure.
 \item To reproduce the straightness of the iso-RM contours, a two-dimensional field 
       structure is needed. In particular, a two-dimensional
       draped field, whose lines are geometrically described
       by a family of ellipses, and associated with compression,
       reproduces the RM bands routinely for any inclination of the
       sources to the line-of-sight (Sec.~\ref{2d}).
       Moreover, it might explain the high RM amplitude and low depolarization observed within
       the bands.
\item  The invariance of the magnetic field along the axis perpendicular to 
       the forward expansion of the lobe suggests that
       the physical process responsible for the draping and stretching of the magnetic 
       field must act on scales larger than the lobe itself in this direction.
       We cannot yet constrain the scale size along the line of sight.
\item In order to create RM bands with multiple reversals, more complex
      field geometries such as two-dimensional eddies are needed
      (Section~\ref{helic}).

\item  We have interpreted the observed RM's as due to two magnetic field components: 
       one draped around the radio lobes to produce the RM bands, the other  
       turbulent, spread throughout the 
       surrounding medium, unaffected by the radio source 
       and responsible for the isotropic and random RM fluctuations (Section~\ref{iso}).
       We tested this model for 0206+35, assuming a typical variation of 
       field strength with radius in the group atmosphere,
       and found that a magnetic field with central strength 
       of $B_0 = 1.8\,\mu$G reproduced the RM range quite well in both lobes.
\item  We have suggested two reasons for the low rate of detection of bands in
       published RM images: our line of sight will only intercept a draped field
       structure in a minority of cases and rotation by a foreground turbulent field with
       significant power on large scales may mask any banded RM structure.
\end{enumerate} 

Our results therefore suggest a more complex picture of the magnetoionic
environments of radio galaxies than was apparent from earlier work. We find
three distinct types of magnetic-field structure: an isotropic component with
large-scale fluctuations, plausibly associated with the undisturbed
intergalactic medium; a well-ordered field draped around the leading edges of
the radio lobes and a field with small-scale fluctuations in the shells of
compressed gas surrounding the inner lobes, perhaps associated with a mixing
layer.  In addition, we have emphasised that simple compression by the bow
shock should lead to enhanced RM's around the leading edges, but that the
observed patterns depend on the pre-shock field.

MHD simulations should be able to address the formation of anisotropic
magnetic-field structures around radio lobes and to constrain the initial
conditions. In addition, our work raises a number of observational questions,
including the following.
\begin{enumerate}
\item How common are anisotropic RM structures? Do they occur primarily in lobed
  radio galaxies with small axial ratios, consistent with jet-driven expansion
  into an unusually dense surrounding medium? Is
  their frequency qualitatively consistent with the two-dimensional draped-field
  picture?
\item Why do we see bands primarily in sources where the isotropic RM component
  has a flat power spectrum of low amplitude? Is this just because the bands can be
  obscured by large-scale fluctuations, or is there a causal connection? 
\item Are the RM bands suggested in tailed sources such as 3C\,465 and Hydra\,A caused by a
  similar phenomenon (e.g.\ bulk flow of the IGM around the tails)?
\item Is an asymmetry between approaching and receding lobes seen in the banded 
  RM component?  If so, what does that imply about the field structure? 
\item How common are the regions of enhanced depolarization at the edges of
  radio lobes? How strong is the field and what is its structure? Is there
  evidence for the presence of a mixing layer?
\end{enumerate}
It should be possible to address all of these questions using a combination of
observations with the new generation of synthesis arrays (EVLA, e-MERLIN
and LOFAR all have wide-band polarimetric capabilities) and high-resolution
X-ray imaging.

\section*{Acknowledgments}
We thank M. Swain for providing the 
VLA data for 3C\,353, J. Eilek for a FITS image of 3C\,465 and 
J. Croston, A. Finoguenov and J. Googder
for the X-ray images of 3C\,270, M\,84 and 3C\,353, respectively.
We are also grateful to A. Shukurov, J. St\"ockl and
the anonymous referee for the valuable comments.

\bsp

\label{lastpage}

\end{document}